\documentclass[a4paper, oneside, 12pt]{Thesis}  % Use the "Thesis" style, based on the ECS Thesis style by Steve Gunn
\graphicspath{{Figures/}}  % Location of the graphics files (set up for graphics to be in PDF format)

\usepackage[square, numbers, comma, sort&compress]{natbib}  % Use the "Natbib" style for the references in the Bibliography
\usepackage{verbatim}  % Needed for the "comment" environment to make LaTeX comments
\usepackage{vector}  % Allows "\bvec{}" and "\buvec{}" for "blackboard" style bold vectors in maths
\usepackage{float}  
\hypersetup{urlcolor=black, colorlinks=black}  % Colours hyperlinks in blue, but this can be distracting if there are many links.

\begin{document}
\frontmatter	  % Begin Roman style (i, ii, iii, iv...) page numbering

% Set up the Title Page
\title  {HTML5 WebSocket protocol and its application to distributed computing}
\authors  {Gabriel L. Muller}
\addresses  {\groupname\\\DEPTNAME\\\univname\\\coursename}  % Do not change this here, instead these must be set in the "Thesis.cls" file, please look through it instead
\date       {\today}
\subject    {}

\maketitle
\makesecondtitle
%% ----------------------------------------------------------------

\setstretch{1.3}  % It is better to have smaller font and larger line spacing than the other way round

% Define the page headers using the FancyHdr package and set up for one-sided printing
\fancyhead{}  % Clears all page headers and footers
\rhead{\thepage}  % Sets the right side header to show the page number
\lhead{}  % Clears the left side page header

\pagestyle{fancy}  % Finally, use the "fancy" page style to implement the FancyHdr headers

%% ----------------------------------------------------------------
% Declaration Page required for the Thesis, your institution may give you a different text to place here
\Declaration{

I, Gabriel L. Muller, declare that this thesis titled, HTML5 WebSocket protocol and its application to distributed computing and the work presented in it are my own. I confirm that:

\begin{itemize} 
\item[\tiny{$\blacksquare$}] This work was done wholly or mainly while in candidature for a research degree at this University.
 
\item[\tiny{$\blacksquare$}] Where any part of this thesis has previously been submitted for a degree or any other qualification at this University or any other institution, this has been clearly stated.
 
\item[\tiny{$\blacksquare$}] Where I have consulted the published work of others, this is always clearly attributed.
 
\item[\tiny{$\blacksquare$}] Where I have quoted from the work of others, the source is always given. With the exception of such quotations, this thesis is entirely my own work.
 
\item[\tiny{$\blacksquare$}] I have acknowledged all main sources of help.
 
\item[\tiny{$\blacksquare$}] Where the thesis is based on work done by myself jointly with others, I have made clear exactly what was done by others and what I have contributed myself.
\end{itemize}

%Signed:\\
%\rule[1em]{25em}{0.5pt}  % This prints a line for the signature
 
%Date:\\
%\rule[1em]{25em}{0.5pt}  % This prints a line to write the date

}
\clearpage  % Declaration ended, now start a new page

%% ----------------------------------------------------------------
% The "Funny Quote Page"
%\pagestyle{empty}  % No headers or footers for the following pages

%\null\vfill
% Now comes the "Funny Quote", written in italics
%\textit{``Write a funny quote here.''}

%\begin{flushright}
%If the quote is taken from someone, their name goes here
%\end{flushright}

%\vfill\vfill\vfill\vfill\vfill\vfill\null
%\clearpage  % Funny Quote page ended, start a new page
%% ----------------------------------------------------------------

% The Abstract Page
\chapter{Abstract}

HTML5 WebSocket protocol brings real time communication in web browsers to a
new level. Daily, new products are designed to stay permanently connected to
the web. WebSocket is the technology enabling this revolution. WebSockets are
supported by all current browsers, but it is still a new technology in constant
evolution.

WebSockets are slowly replacing older client-server communication
technologies. As opposed to comet-like technologies WebSockets' remarkable
performances is a result of the protocol's fully duplex nature and because it
doesn't rely on HTTP communications.

To begin with this paper studies the WebSocket protocol and different 
WebSocket servers implementations. This first theoretic part focuses more 
deeply on heterogeneous implementations and OpenCL. The second part is
a benchmark of a new promising library. 

The real-time engine used for testing purposes is SocketCluster. SocketCluster
provides a highly scalable WebSocket server that makes use of all available cpu
cores on an instance. The scope of this work is reduced to vertical scaling of 
SocketCluster.

\clearpage  % Abstract ended, start a new page
%% ----------------------------------------------------------------

\setstretch{1.3}  % Reset the line-spacing to 1.3 for body text (if it has changed)

% The Acknowledgements page, for thanking everyone
\chapter{Acknowledgements}

I wish to thank my supervisor Mark Stillwell for accepting to work with me on
this project. His positive suggestions and willingness to repeadtedly meet and
discuss the progress of this thesis have been invaluable.

Naturally, I am also grateful for the continuous support of my family and of my
housemates. Always available and willing to provide either advises or
distractions.

Lastly I would like to thank my colleage L\'eo Unbekandt. Not only did he help
me in the very begenning when I was looking for a subject, but also later on
through the whole thesis. More importantly our daily lunch conversations
confirmed and raised my interest in computer science. Thanks L\'eo.

\clearpage  % End of the Acknowledgements
%% ----------------------------------------------------------------

\pagestyle{fancy}  %The page style headers have been "empty" all this time, now use the "fancy" headers as defined before to bring them back

%% ----------------------------------------------------------------
\lhead{\emph{Contents}}  % Set the left side page header to "Contents"
\tableofcontents  % Write out the Table of Contents

%% ----------------------------------------------------------------
\lhead{\emph{List of Figures}}  % Set the left side page header to "List if Figures"
\listoffigures  % Write out the List of Figures

%% ----------------------------------------------------------------
% \lhead{\emph{List of Tables}}  % Set the left side page header to "List of Tables"
% \listoftables  % Write out the List of Tables

%% ----------------------------------------------------------------
\setstretch{1.5}  % Set the line spacing to 1.5, this makes the following tables easier to read
\clearpage  % Start a new page
\lhead{\emph{Abbreviations}}  % Set the left side page header to "Abbreviations"
\listofsymbols{ll}  
{
\textbf{RFC} & Request For Comment \\ 
\textbf{HTTP} & HyperText Transfert Protocol \\ 
\textbf{HTML} & HyperText Markup Language \\ 
\textbf{TCP} & Transmission control protocol \\ 
\textbf{IP} & Internet Protocol \\ 
\textbf{UDP} & User Datagram Protocol  \\ 
\textbf{OpenCL} & Open Computing Language \\ 
\textbf{OpenGL} & Open Graphic Library \\ 
\textbf{API} & Application Programming Interface \\ 
\textbf{GPU} & Graphic Processing Unit \\ 
\textbf{GPGPU} & General Purpose computation on GPU \\ 
\textbf{CPU} & Computing Processor Unit\\ 
\textbf{SIMD} & Single Instruction Multiple Data \\ 
\textbf{MIME} & Multi purpose Internet Mail Extension \\ 
\textbf{DSP} & Digital Signal Processing \\ 
\textbf{VCL} & Virtual open Computing Language \\ 
\textbf{TFLOPS} & Tera FLoating Operation Per Seconds \\ 

}
 
%% ----------------------------------------------------------------
% End of the pre-able, contents and lists of things
% Begin the Dedication page

\setstretch{1.3}  % Return the line spacing back to 1.3

\pagestyle{empty}  % Page style needs to be empty for this page
%\dedicatory{For/Dedicated to/To my\ldots}

\addtocontents{toc}{\vspace{2em}}  % Add a gap in the Contents, for aesthetics

%% ----------------------------------------------------------------
\mainmatter	  % Begin normal, numeric (1,2,3...) page numbering
\pagestyle{fancy}  % Return the page headers back to the "fancy" style

% \label{Chapter1}
\chapter*{Introduction}
\lhead{\emph{Introduction}}

\textbf{Problem statement}

WebSockets are implemented in a wide range of applications. As a result, a lot
of different languages and specific libraries have been specifically developed
for WebSockets.

The first part of this paper is an introduction to the WebSocket protocol and
on the different implementation options when building a WebSocket cluster. In a
second part, it focuses on the new real-time engine: SocketCluster and makes a
benchmark of this library.

\textbf{Thesis structure}

The first chapter is a literature review. The goal is to inform the reader
about the WebSocket protocol and to go through the different WebSocket
implementations. Therefore studying scalability and heterogeneous
implementations.

The second chapter is an introduction to the experiments. It is dedicated to the
design and the implementation of the infrastructure used later on. It mostly 
introduces the benchmark library used.

The experiment chapter is a comprehensive benchmark of SocketCluster. It 
compares the performances to a classic engine.io implementation and also
studies the limitations of the library.

To finish the last part concludes this thesis and suggest future work on 
SocketCluster.

\addtotoc{Introduction}

\chapter{Literature review} 
\label{Chapter1} 
\lhead{Chapter 1. \emph{Literature review}}

This chapter is an introduction to the WebSocket protocol.  It begins with a
section which sums up client-server communication techniques.  The second
section is an in depth study of WebSockets. The third section is about 
WebSockets servers' implementations, the last is about scalability. 

\section{Client-server communications}

This section studies the evolution of client-server communication, beginning
with the page by page model until current technologies. However as an
introduction, the first part is about HTPP which is the foundation of
client-server communication.

\subsection{HTTP protocol}

The HTTP protocol is a request/response protocol defined in the request for
comment (RFC) \cite{Reference1} as follows:

\begin{verbatim} 
A client sends a request to the server in the form of a request 
method, URI, and protocol version, followed by a MIME-like 
message containing request modifiers, client information, and 
possible body content over a connection with a server. The 
server responds with a status line, including the message's
protocol version and a success or error code, followed by a 
MIME-like message containing server information, entity meta 
information, and possible entity-body content.  
\end{verbatim}

Because HTTP was not designed for real time communication several workarounds
have been developed over the years to overcome the so called page by page
model. These techniques are Explained in details in Eliot Step master thesis
\citep{Reference2}.

\subsection{Page by page model}

Since HTTP's release in 1991, client-server communications have undergone
continuous upgrades. In the early nineties, most web pages were static. As a
consequence, the communication between client and server  were rather limited.
Typically, the client would send occasional requests to the server. The server
would then answer, but all communication would stop there until a new event was
triggered by the user.

\begin{figure}[H]
\centering
\includegraphics[width=0.9\textwidth]{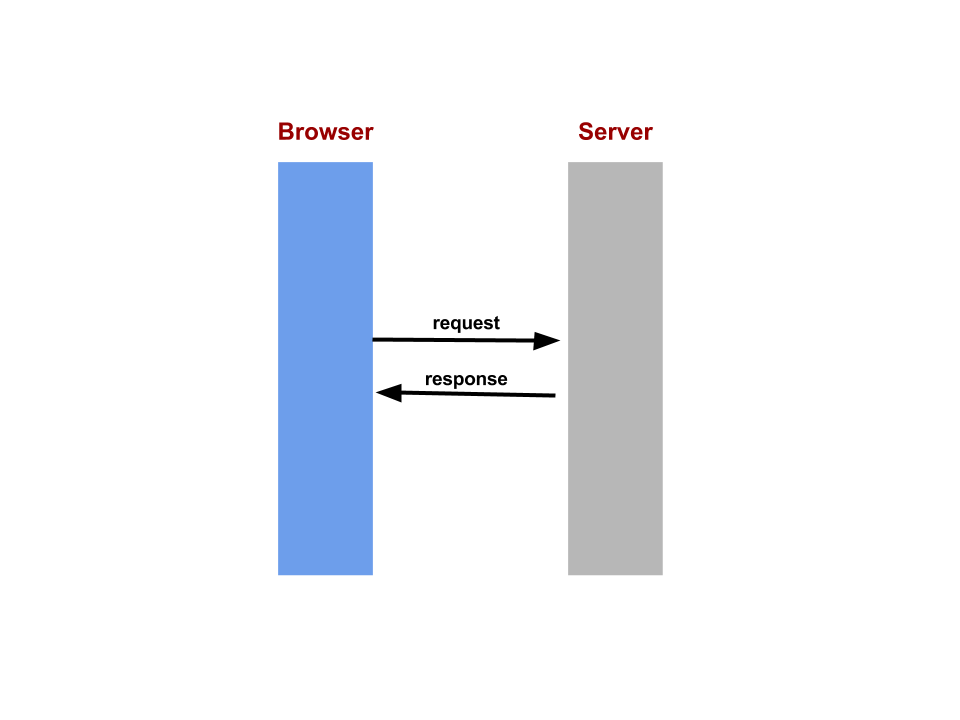}
\caption[Client-server communication]{Client-server communication}
\label{fig:client_server_communication}
\end{figure}

The notion of dynamic web appeared in 2005 with the introduction of technologies
like Comet. Peter Lubbers describes it as the Headache 2.0 in his article
\texttt{"A quantum leap in scalability for the web"} \citep{Reference32}.

\subsection{Polling}

Polling was the first attempt toward real-time communication. Instead of waiting
for the client to manually ask for a page update, the browser would send regular
HTTP GET requests to the server. This technique could be efficient if the exact
interval of update on the server side was known.

\begin{figure}[H]
\centering
\includegraphics[width=\textwidth]{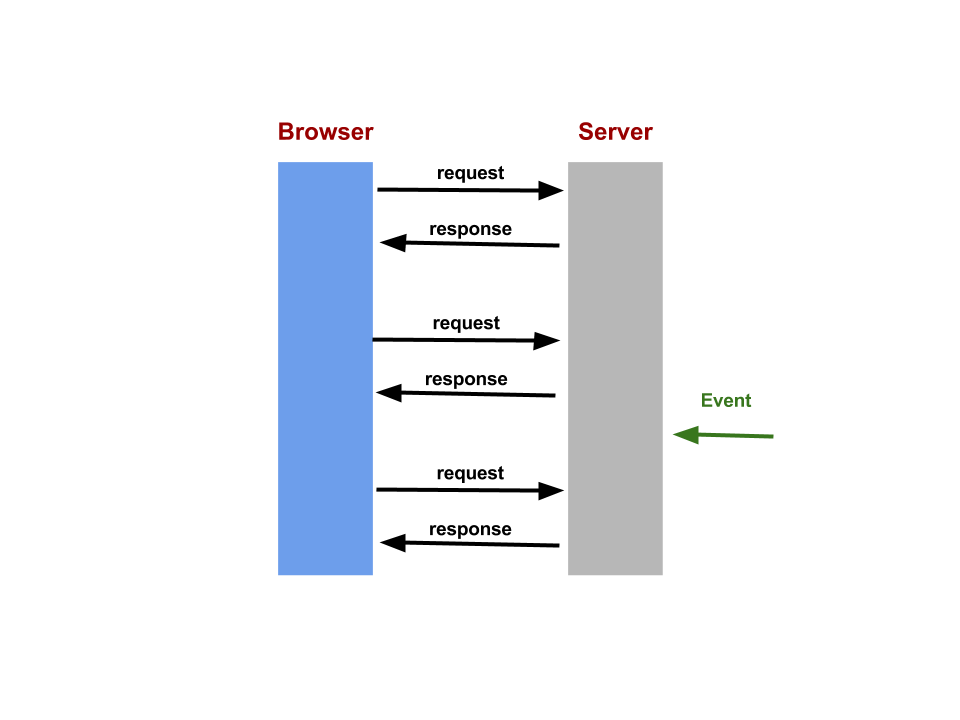}
\caption[Polling]{polling}
\label{fig:polling}
\end{figure}

However real time information is unpredictable and in high updates rate
situation like stock prices, news reports or tickets sales the
response could be stale by the time the browser renders the page
\citep{Reference32}.

Also in low updates rate situation even if no data is available, the server
will send an empty response. This would result in a large amount of unnecessary
connections being established, which over time and with the clients increase
would lead to decreased overall network throughput \citep{Reference2}.

\subsection{Long polling}

Long polling is based on Comet technologies and is a slight step further toward
server sent events and real time communication. Comet began to be popular in web
browser around 2007, it is a family of web techniques that allows the server to
hold an HTTP request open for prolonged periods of time.

\begin{figure}[H]
\centering
\includegraphics[width=0.9\textwidth]{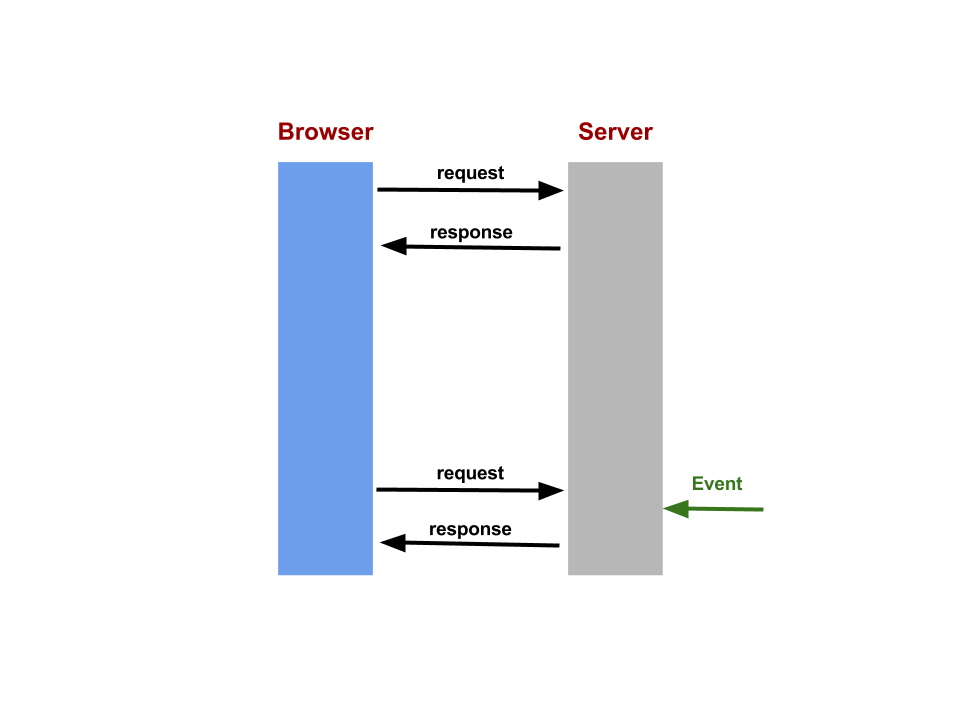}
\caption[Long polling]{Long polling}
\label{fig:long_polling}
\end{figure}

Long-polling is similar to polling, except that the server keeps the HTTP
request open if data is not immediately available. The server determines how
long to keep the request open, request also known as a \texttt{hanging GET}. If
new data is received within the time interval, a response containing the data
is sent to the client and the connection is closed. If new data is not received
within the time period, the server will respond with a notification to
terminate the open request and close the connection. After the client browser
receives the response, it will create another  request to handle the next
event, therefore always keeping a new long-polling request open for new events.
This results in the server constantly responding with new data as soon as it is
made available \citep{Reference2}.

However, in situations with high-message volume, long- polling does not provide
increased performance benefits over regular polling. Performance could actually
be decreased if long-polling requests turn into continuous, unthrottled loops
of regular polling requests.

\subsection{Streaming}

Streaming is based on a persistent HTTP connection. The communication still
begins with a request from the browser, the difference is in the response. The
server never signals the browser its message is finished. This way the
connection is kept open and ready to deliver further data \citep{Reference2}.

\begin{figure}[H]
\centering
\includegraphics[width=0.9\textwidth]{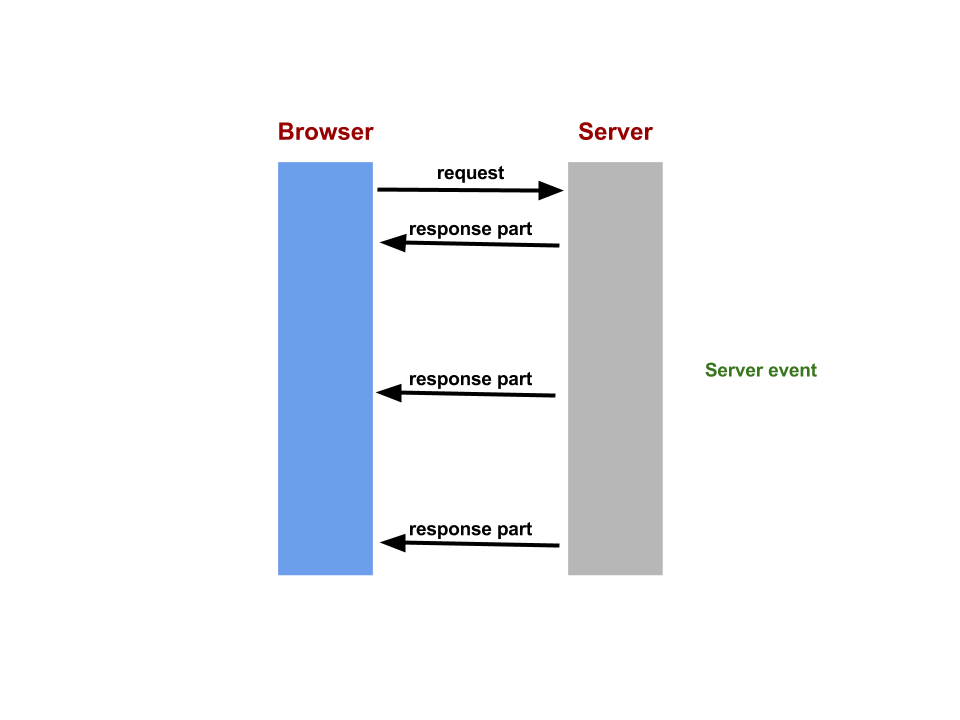}
\caption[Streaming]{Streaming}
\label{fig:streaming}
\end{figure}

If it wasn't for proxies, streaming would be perfectly adapted for real time
communication. Because streaming is done over HTTP, proxy server may choose to
buffer server responses and thus increasing greatly the latency of the message
delivery. Therefore in case a proxy is detected most Comet-like solution fall
back to long polling \citep{Reference2}.

\subsection{Current technologies in browsers}

At the moment, comet technologies are still the most popular way of
communication between browsers and servers. Techniques has been improved to
the point where it perfectly fakes server sent event. Comet technologies can be
seen as a wonderful hack to reach real time communication. However little can
be done to improve the latency. Comet technologies revolve around HTTP and
carry its overhead.

The total overhead from the HTTP request and response header is at least 871
bytes without containing any data. In comparison, a small payload is 20 bytes.
Contemporary application like on-line games can not be built on a technology
wasting resources equivalent to 40 messages every time information is 
exchanged \citep{Reference2}. Therefore a brand new protocol has been
developed: WebSocket.

\section{WebSocket protocol}

The creation of the WebSocket protocol marks the beginning of the Living web.
It is often referred to as the first major upgrade in the history of web
communications. As the Web itself originally did, WebSocket enables entirely
new kinds of applications. Daily, new products are designed to stay permanently
connected to the web. Websocket is the language enabling this revolution.

This section is a study of the WebSocket protocol. Firstly it defines the
protocol. Secondly it studies how to establish a WebSocket connection.
Afterwards it goes on with an in depth study of WebSockets' transport layer and
frame anatomies. Lastly it provides a brief discussion of WebSockets'
interaction with proxies.

\subsection{Definition}

The official Request For Comments \citep{Reference12} (RFC) describes the
WebSocket protocol as follows:

\begin{verbatim}
The WebSocket Protocol enables two-way communication between a 
client running untrusted code in a controlled environment to a 
remote host that has opted-in to communications from that code.
The security model used for this is the origin-based security 
model commonly used by web browsers. The protocol consists of an
opening handshake followed by basic message framing, layered over
TCP. The goal of this technology is to provide a mechanism for
browser-based applications that need two-way communication with
servers that does not rely on opening multiple HTTP connections.
\end{verbatim}

To Initiate a WebSocket communication, first a HTTP handshake needs to be done.

\subsection{The WebSocket handshake}

The Websocket protocol was to be released in an already existing web
infrastructure. Therefore it has been designed to be backward-compatible. Before
a Websocket communication can start, a HTTP connection must be initiated. The
browser sends an Upgrade header to the server to inform him he wants to start a
WebSocket connection. Switching from the HTTP protocol to the WebSocket
protocol is referred to as a handshake \citep{Reference12}.

\begin{verbatim}
GET ws://websocket.example.com/ HTTP/1.1
Origin: http://example.com
Connection: Upgrade
Host: websocket.example.com
Upgrade: websocket
\end{verbatim}

If the server supports the WebSocket protocol, it sends the following header in
response.

\begin{verbatim}
HTTP/1.1 101 WebSocket Protocol Handshake
Date: Wed, 5 May 2014 04:04:22 GMT
Connection: Upgrade
Upgrade: WebSocket
\end{verbatim}

After the completion of the handshake the WebSocket connection is active and
either the client or the server can send data. The data is contained in frames,
each frame is pre-fixed with a 4-12 bytes to ensure the message can be
reconstructed. 

Once the server and the browser have agreed on beginning a WebSocket
communication. A first request is made to begin an ethernet communication
followed by a request to make an TCP / IP communication.

\subsection{Transport layer protocol}

The internet is based on two transport layer protocols, the User Datagram
Protocol (UDP) and the Transmission Control Protocol (TCP). Both use the
network layer service provided by the internet protocol (IP). 

\textbf{TCP}

TCP is a reliable transmission protocol. The data is buffered byte by byte in
segments and transmitted according to specific timers. This flow control ensure
the consistency of the data. TCP is said to be a stream oriented because the
data is sent in independent segments.

\textbf{UDP}

UDP is unreliable but fast. The protocol offers no guaranty the data will be
delivered in its integrality nor duplicated. It works on a best effort strategy
with no flow control. Each segments are received independently, it is a message
oriented protocol.

Websocket is build over TCP because of its reliability. Browser enabled games
are the perfect example of WebSockets' use cases. They require low latency and
have a high rate of update. To achieve low latency, the communication protocol
must make sure not to drop any packets. Otherwise, the exchange takes two times
longer.

As can be inferred from the 2 previous subsections, the websockets protocol
relies on a few other protocols. Namely HTTP to initialize the communication ,
ethernet, TCP/IP and finally TLS in case a secure connections is required.  The
next subsections studies the influence this protocols have in the anatomy of
WebSockets frame.

\subsection{The WebSocket frame anatomy}

\begin{figure}[H]
\centering
\includegraphics[width=0.6\textwidth]{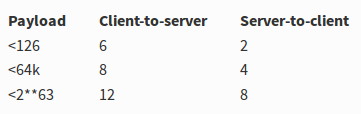}
\caption[Frame overhead]{Frame overhead \citep{Reference30}}
\label{fig:frameOverhead}
\end{figure}

The study conducted by Tobias Oberstein \citep{Reference30} looks into the
overheads of websockets. As a matter of fact the overhead induced purely by
WebSockets is extremely low. As can be seen in the figure
\ref{fig:frameOverhead}, depending on the size of the payload the overhead
varies between 8 and 20 bytes.

\begin{figure}[H]
\centering
\includegraphics[width=\textwidth]{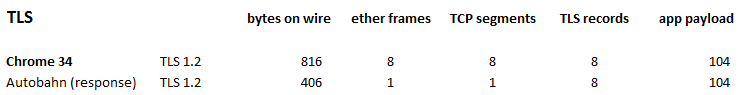}
\caption[TLS overhead]{TLS overhead \citep{Reference30}}
\label{fig:tlsOverhead}
\end{figure}

\begin{figure}[H]
\centering
\includegraphics[width=\textwidth]{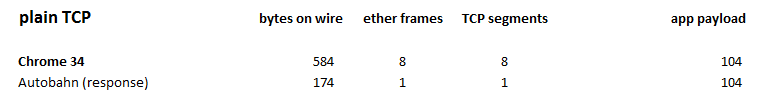}
\caption[TCP overhead]{TCP overhead \citep{Reference30}}
\label{fig:tcpOverhead}
\end{figure}

However, as pointed out in the article efficiency is lost on protocols of other layers
required for WebSocket's functionment. Figure \ref{fig:tlsOverhead} and
\ref{fig:tcpOverhead} respectively show the overhead induced by pure TCP/IP and
TLS protocols.

In this example, the payloads \texttt{Hello world} is only thirteen bytes. In
comparison ethernet, TCP/IP and TLS protocols each use height bytes. The
conclusion of this article is to warn programmers about the size of the
payloads so that all the protocols revolving around WebSockets don't dwarf the
overhead of the WebSocket protocol itself. In case small payloads can not be
avoided a possible solution is to serialize the messages in order to batch them
in one single WebSocket message.

\begin{figure}[H]
\centering
\includegraphics[width=\textwidth]{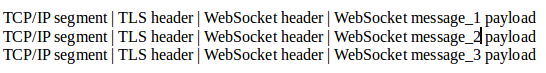}
\caption[Websocket messages sent individually]{WebSocket messages sent individually \citep{Reference30}}
\label{fig:separateWebsocket}
\end{figure}

\begin{figure}[H]
\centering
\includegraphics[width=0.7\textwidth]{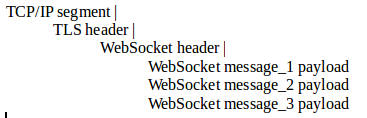}
\caption[Batched WebSocket messages]{Batched WebSocket messages \citep{Reference30}}
\label{fig:batched_websocket}
\end{figure}

So instead of sending the each messages using the WebSocket protocol like shown
in figure \ref{fig:separateWebsocket}. The individual messages are put in a
queue and batched in a single Websocket message like in figure
\ref{fig:batched_websocket}.

\vspace{10 mm}

Nevertheless, WebSockets carry way less overhead then comet technologies do.
Another advantage of WebSocket its interaction with proxies.

\subsection{Proxies}

Proxy servers are set up between a private network and the Internet. They act
like an intermediary providing content caching, security and content
filtering.

When a Websocket server detects the presence of a proxy server, it
automatically sets up a tunnel to pass through the proxy. The tunnel is
established by issuing an HTTP CONNECT statement to the proxy server, which
requests for the proxy server to open a TCP/IP connection to a specific host
and port. Once the tunnel is set up, communication can flow unimpeded through
the proxy.

To sum up compared to comet technologies, WebSockets are: 

\begin{itemize}
    \item As reliable, because they are also built over TCP
    \item Way faster, because they don't carry the overhead of HTTP
    \item Behaving better in presence of proxies
    \item Fully bidirectional
\end{itemize}

The nexts sections of this chapter are dedicated to the implementation of 
WebSockets servers.

\section{Implementation}

As in any project, in order to avoid future technical problems, it is better to
first study similar projects. The goal of this implementation study is to find a
suitable language and possibly a good library to run the experiment.

\subsection{WebSocket server implementation}

In order to narrow the library study, first a language needs to be selected.

\textbf{Language Selection}

Choosing a language for a project is often a compromise between the programmers
development background and the necessity of the application. Furthermore,
WebSocket servers can be developed in almost any languages.

This subsection does not aim at giving a comprehensive comparison of all
existing WebSocket friendly languages. Node.js seems to be the perfect
environment for this study, therefore other languages will deliberately be left
out and the focus will be on explaining why Node.js is appropriate.

Node.js was specially invented to create real-time websites with push
capabilities \citep{Reference35}. Most languages run parallel tasks by using
threads but threads are memory expensive. Node.js is fundamentally different,
it runs as a single non-blocking and event-driven loop by using asynchronous
call back loops \citep{Reference37}. For this reasons, compared to other
languages, Node.js performs significantly better in highly concurrent
environment.

Node.js has many real-time engines. The next step is to carefully make a choice
between ws, Socket.io and Engine.io.  

\textbf{WebSocket implementation selection}

Deniz Ozger article's for medium.com \citep{Reference36} is a comprehensive study
of node.js real-time engines.

Ws is a pure WebSocket implementation, therefore it is appropriate for testing
purposes but seldom used in real life projects.  The main drawback is the
communication may not work in case the browser does not support WebSockets.

Socket.io has some appreciable features namely its connection procedure. First
it tries to connect to a server via WebSocket, in case it fails it downgrades
until it finds a suitable protocol. Moreover it tries to reconnect sockets
when connections fail. 

Engine.io is a lower library of Socket.io. The connection procedure is the
opposite to Socket.io though. It first establishes a long polling connection
and only later tries to upgrade it to a better transport protocol. Therefore it
is more reliable because it establishes less connection.

In conclusion, Node.js and its real-time library engine.io seems to best choice
for our study. However better performance could be reached using an
heterogeneous implementation.

\subsection{Heterogeneous implementation with OpenCL}

As suggest John Stone paper's title \texttt{"OpenCL: A parallel programming standard
for heterogeneous computing systems"} \citep{Reference5} OpenCL is unanimously
considered as the reference for heterogenous computing.

Historically, the first technology to take advantage of the massive parallel
nature of GPUs was Open Graphic Library (OpenGL). OpenGL is an application
programming interface (API) for rendering 2D and 3D vector graphics. Through
the insertion of little pieces of C-like codes in shader, developers soon
realized graphic processing units (GPUs) could also be used for general
programming. This became known as General Purpose computation on GPUs (GPGPU)
\citep{Reference5}.

However, shaders can only be modified so much. As the need for more complex
applications arose, Apple proposed the Khronos Group to develop a more general
framework: OpenCL. OpenCL is a low-level API accelerating applications with
task-parallel or data-parallel computations in a heterogeneous computing
environment. Indeed OpenCL not only allows the usage of CPUs but also any
processing devices like GPUs, DSPs, accelerators and so on \citep{Reference5}.
If generally, on desktops the diversity of processing devices is quite low, as
opposed to mobile devices. Embedded systems for real-time multimedia journal
published a paper \citep{Reference3} high lining the advantages of using OpenCl
in mobile browser.

OpenCL doesn't guarantee a particular kernel will achieve peak performance on
different architectures. The nature of the underlying hardware may induce
different programming strategies. Multi-core CPU architecture is definitely the
more popular. But the recent specification published by Khronos to take GPU
computing to the web is bound to raise programmers interest toward GPUs
architecture \citep{Reference30}. 

\textbf{CPUs architecture}

Modern CPUs are typically composed of a few high-frequency processor cores.
CPUs perform well for a wide variety of applications, but they are optimal for
latency sensitive workloads with minimal parallelism. However, to increase
performance during arithmetic and multimedia workloads,  many CPUs also
incorporate small scale use of single instruction multiple data (SIMD).

\textbf{GPUs architecture}

Contemporary GPUs are composed of hundreds of processing units running at low
frequency. 

As a result GPUs are able to execute tens of thousands of threads. It is this
ability which makes them so much more effective then CPUs in a highly parallel
environment. Some research even claim a speedup in the order of 200x over
JavaScript. \citep{Reference3}

The GPU processing units are typically organized in SIMD clusters controlled by
single instruction decoders, with shared access to fast on-chip caches and
shared memories. Massively parallel arithmetic-heavy hardware design enables
GPUs to achieve single-precision floating point arithmetic rates approaching 2
trillions of instructions per second (TFLOPS). \citep{Reference5}

Although GPUs are powerful computing devices, currently they still often
require to be management by a host CPU. Fortunately OpenCL is designed to be used in
heterogeneous environment. It abstracts CPUs and GPUs as “compute devices”.
This way, applications can query device attributes to determine the
properties of the available compute units and memory systems.
\citep{Reference5}

All the same, even if OpenCL's API hides the hardest part of parallel
programming a good understanding of the underlying memory model leads to more
efficient coding. Along with general advises on how to build an OpenCL
cluster, details about the memory model are given in the following paper:
\citep{Reference4}.

\textbf{Platform model}

CPU and GPU are called “compute devices”. A single host regroups one or more
compute devices and has its own memory. Each compute device is composed of one
or more cores also called “compute units”. Each compute unit has its own memory
and is divided into one or more SIMD threads or “processing elements” with its
own memory. \citep{Reference4}

\begin{figure}[H] \centering
\includegraphics[width=\textwidth]{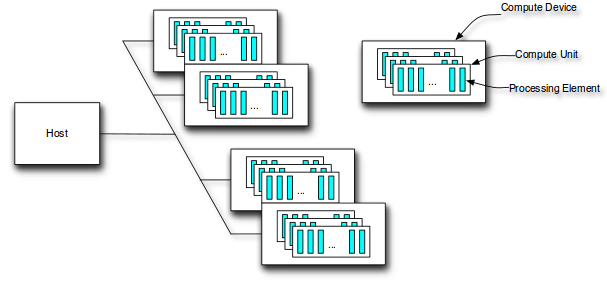}
\caption[Platform model]{Platform model \citep{Reference4}}
\label{fig:plateform_model} \end{figure}

\textbf{Memory model}

OpenCL defines 4 types of memory spaces within a compute device. A large
high-latency “global” memory corresponding to the device RAM. This is a none
cached memory  where the data is stored and is available to all items. A small
low-latency read-only “constant” memory which is cached. A shared “local”
memory accessible from multiple processing elements within the same compute
unit and a “private” memory accessible within each processing element. This
last type of memory is very fast and is the register of the items
\citep{Reference4}.

\begin{figure}[H] \centering
  \includegraphics[width=0.6\textwidth]{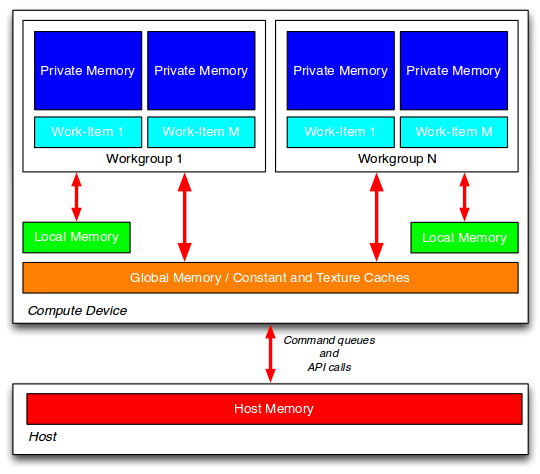}
  \caption[Memory model]{Memory model \citep{Reference4}}
  \label{fig:memory_model} 
\end{figure}

In conclusion, OpenCL provides a fairly easy way to write parallel code but to
reach an optimal performance / memory access trade off programmers must choose
carefully in where to save their variables in memory space.

\textbf{Global and local IDs}

Finally, at an even lower level, work-items are scheduled in work–groups.
This is the smallest unit of parallelism on a device. Individual work-items in
a work–group  start together at the same program address, but they have their
own address counter and register state and are therefore free to branch and
execute independently \citep{Reference4}.

\begin{figure}[H] \centering
  \includegraphics[width=\textwidth]{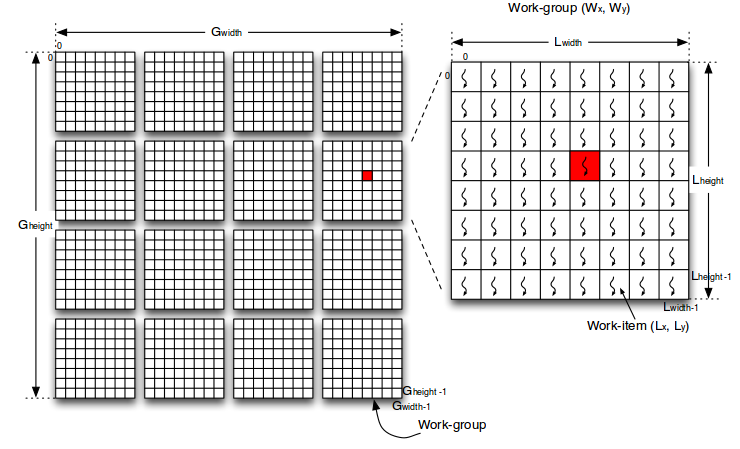} 
  \caption[Work - group]{Work Group \citep{Reference4}} 
  \label{fig:id} 
\end{figure}

On a CPU, operating systems often swap two threads on and off execution
channels. Threads (cores ) are generally heavyweight entities and those context
switches are therefore expensive. By comparison, threads on a GPU ( work-items
) are extremely lightweight entities. Furthermore in GPUs, registers are
allocated to active threads only. Once threads are complete, its resources are
de-allocated.  Thus no swapping of registers or state occurs between GPU
threads. \citep{Reference4}

It can be deduced from this section that the underlying memory model, OpenCL is a
fairly low-level API. In fact, the programming language used is a derivate of
the C language based on C99. A language web developers will most likely be
unfamiliar with. Khronos anticipated this and developed the web computing
language (WebCL).

\subsection{WebCL}

WebGL and WebCL are JavaScript APIs over OpenGL and OpenCL's API. This allows
web developers to create application in an environment they are used to.

In the first place, OpenCL was developed because of web browsers' increasing
need for more computational power. A necessity which arose from heavy 3D
graphics applications such as on-line games and augmented reality. However,
OpenCL doesn’t provide any rendering capability, it only processes huge amounts
of data. That is why OpenCL was designed for inter-operation with OpenGL.
WebCL/WebGL interoperability builds on that available for OpenCL/OpenGL. WebCL
provides an API for safely sharing buffers with OpenCL. This buffer is inside
the GPU which avoids the back and forth copy of data when switching between
OpenGL and OpenCL processes. Further precision about the interoperability are
discussed in this paper: \citep{Reference6}.

GPU computing is quite a new notion. But it is a fast evolving field of
research. Single GPUs are not enough anymore, the trend is moving towards GPU
clusters.

\subsection{GPU clusters}

Most OpenCL applications can utilize only devices of the hosting computer. In
order to run an application on a cluster, the program needs to be split to take
advantage of all devices. Virtual OpenCL (VirtualCL) is a wrapper for OpenCL.
It provides a platform where all the cluster devices are seen as if located on
the same hosting node. Basically, the user starts the application on the master
node then VirtualCL transparently runs the kernels of the application on the
worker nodes. Applications written with VirtualCL don't only benefit from the
reduced programming complexity of a single computer, but also from the
availability of shared memory and lower granularity parallelism. Mosix white
paper \citep{Reference7} explains more in depth the VCL's functionment.

OpenCL and VirtualCL are powerful tool to create highly parallel clusters.  But
current implementation with CPUs only already reach a million concurrent
connections \citep{Reference13}. So far there is simply no need for more
powerful clusters.

However, all company don't have access to dual Quad-core Xeon CPUs used in
Kaazing cluster to reach a million concurrent connections. Usual practice is to
build a scalable cluster, to adjust computing power in function of the needs.

\section{Scalability}

The growth of distributed computing has changed the way web application are
designed and implemented. If compared with today standards, applications used
to be deployed so as to say at prototype stage. That is, they were designed to
work on a fixed number of servers and not able to adjust as the user base grows.
As the number of connections increase, the load on the servers rises and thus
the latency grows. Ideally, an application should aim at a stable latency,
otherwise the application can miss behave.

On the server side, the nodes will begin to be overloaded and struggle to
service the client with reasonable response time.

Also, if the servers are overwhelmed they buffer the responses to the clients
and then catch up later on . As a result, the clients can be flooded when the
load goes down. The sudden rush of message can provoke an unexpected behavior
from the servers and can even lead to disconnections.

Nowadays, designing an application without scalability and load balancing in
mind is unimaginable. Historically, the reaction to an overloaded server
has always been to scale up.

\subsection{Scaling up}

Scaling up or vertically basically means upgrading the infrastructure.
Depending on the needs of the application, the processor, the memory, the
storage or the network connectivity can be improved.

Further performance can be gained by dividing tasks. It only requires
identification of the services  running independently or the using message
based communication. Those could then be relocated on different nodes.

The main advantage of scaling vertically is that is does not involve any software
changes and little infrastructure changes. Therefore it is an easy way to
increase performances. However for large applications, scaling up might prove
impossible or at least not economically profitable. In case the infrastructure 
is already equipped with the latest hardware generation, the tiniest
increase in performance will impact greatly the price. For example, a high
range processor offering ten percent more computation power is going to be
many times more expensive. Similarly, a memory upgrade could require replacing
all current modules for higher density ones.

Moreover, scaling up neither answers availability nor uptime concerns. The
system is monolithic and has a single point of failure. Therefore contemporary
project usually scale out and use parallel computing.

\subsection{Scaling out}
				
Scaling out or horizontally, answers most of the problems unsolved by scaling
vertically. In a first approach lets ignore the software complexity.  Scaling
out offers almost unlimited performance increase and at low cost. If the
application is designed to be spread out on multiple nodes, the performance of
an infrastructure can be doubled by simply using twice as many servers. Also it
is fairly easy to add some redundant server to insure uptime. Plus, compared to
scaling up, once the software is developed the costs are linear.

When scaling out, the infrastructure implementation is not as much of a problem as
the code implementation. The expenses are shifted from hardware to development
costs.

\textbf{Code implementation}

Developing a parallel code is quite complicated and all applications can not
be paralyzed. In 1967 Gene M. Amdhal defined the so called Amdahl's law which
is still used today to define the maximum to expect when parallelizing a code
\citep{Reference10}. 

Each software can be divided in two separate parts, the parallel part and the
sequential part. Parallel computing does not improve the sequential part. If a
the code is mainly sequential, then increasing the number of processors will
only cause the parallel part to finish first and stay idle waiting for the
sequential part to finish.

Assuming P is the portion of a program that can be parallelized and 1 - P  is
the portion that remains serial, then the maximum speedup that can be achieved
using N processors is: 

$$speedup(N) = \frac{1}{(1-P) + \frac{P}{N}} $$

If 70\% of the program can be run in parallel (P = 0.7) the maximum expected
speedup with 4 processors would be:

$$speedup(N) = \frac{1}{(1-0.7) + \frac{0.7}{N}}$$

$$speedup(4) = 2.1$$

When the number of processors reaches a certain point, the speed up will be:

$$\lim\limits_{N \to \infty} speedup(N)= \frac{1}{1-P} = 3.3$$

Nathan T. Hayes's paper for Sunfish Studio \citep{Reference8} studies how
parallel computing can profit the motion picture Industry. The following chart
presents the maximum speedup which can be expected from an application in
function of the percentage of parallel code in the programme.

\begin{figure}[H] 
  \centering
  \includegraphics[width=0.5\textwidth]{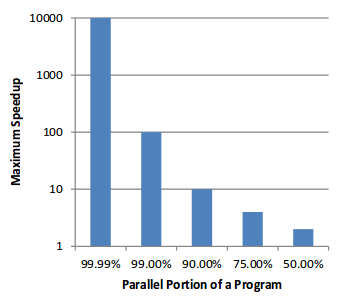}
  \caption[Amdahl law]{Amdahl law \citep{Reference8}} 
  \label{fig:amdahl} 
\end{figure}

The figure speaks for itself, to envisage parallel computing, the portion of
parallel code must be very high.

However, Amdahl's law is based on assumption which are hardly verified in
pratique. Following are summed up reasons not to give to much importance to
Amdahl's law \citep{Reference34}:

\begin{itemize} 
  \item The number of threads is not always equivalent to the number of processors.  
  \item The parallel portion does not have a perfect speedup. Computation power is used
    for communication between processus. Also some resources like caches and bandwidth
    have to be shared across all the processors.  
  \item Allocating, deallocating and switching threads introduces overhead, overhead grows
    linearly with the number of thread.  
  \item Even an optimized code will not have perfectly synchronised threads, at some point
    some processus will have to wait for others to finish.	
\end{itemize}

Amdahl's law has long been used as an argument against massively parallel
processing. In 1988 Gustafson law came as an alternative to Amdahl's law to
estimate the speedup. In both law, the sequential portion of the problem is
supposed to stay constant. But in Gustafson's law the overall problem size
grows proportionally to the number of cores. As a result, Gustafson's gives
slightly different results to Amdahl's and encourage the use of parallel
computing.

However later studies tends to contest the legitimacy of both laws. Yuan Shi's
paper \citep{Reference9} even proves both theories are but two different
interpretations of the same law. He concludes his study by saying these laws
are too minimalist and what computer scientist really need is a practical
engineering tool that can help the community to identify performance critical
factors.

\textbf{Infrastructure implementation}

Beside coding complication, scaling out also brings infrastructure changes.
A third party must be in command of all servers. This master server is also called
load balancer. Its role is to distribute the work evenly between the workers and thus
completely hides the complexity to the user.

\chapter{Design and Implementation} 
\label{Chapter2} 
\lhead{Chapter 2. \emph{Design and implementation}} 

Current research around WebSocket is centered around distributed computing. Either
on CPUs architecture, GPUs architecture or heterogeneous architecture. For the
time being, clustering WebSocket servers is rather difficult and reserved to
researcher or specialized companies. Actually in Node.js, there is hardly any
library to simply and efficiently implement a multi-core server.  Node.js
single thread nature is a double edge sword. On one side it allows more
concurrent connections to be established but it also means special attention
needs to be given to run the code on all the servers cores. SocketCluster is a
brand new real-time engine aiming exactly at that.  At this point of my thesis
I had to make a choice between either the theoretical study of WebSocket
clusters or the benchmarking of SocketCluster. After contacting Jonathan
Gros-Dubois, the creator of SocketCluster, I made up my mind for the latter.
Indeed, SocketCluster being under development the tests carried out so far are
rather sparse. 

\section{SocketCluster library}

As described on the github project  \citep{Reference38}, SocketCluster is a
fast, highly scalable HTTP and WebSocket server. It facilitates the creation of
multi-process real-time application that make use of all CPU cores on a
machine/instance. Therefore removing the limitations of having to run a Node.js
server as a single thread.  SocketCluster's focus is on vertical scaling. If N
is the number of cores available on the server, then SocketCluster is N time
than any comparable single-threaded WebSocket server. Under the hood, the
application deploys itself on all available cores as a cluster of process. The
process can be grouped in three categories: stores, workers and load balancers.

\section{Challenges encountered using SocketCluster}

At first my study of SocketCluster was far from satisfactory. Past a
total of 512 communications channel, new sockets were inexplicably crashing.

\textbf{U-limit}

This comes from a system limit set up on linux operating systems. By default
the maximum number of file that can be sent over tcp is 1024. 

Fortunately, this limit can be increased by appending this line: \texttt{ubuntu
soft nofile "number of file"} in \texttt{/etc/security/limits.conf}

Once this problem was fixed I looked into a benchmark to carry out, Jonathan Gros-Dubois
advised me to focus on concurrent connections tests.

\textbf{C 10K Problem}

The C 10K is an historic challenge issued in 1999 by Dan Kegel. It consist of
reaching 10 000 concurrent client connections. Engineers solved this problem by
fixing operating systems kernel and creating new single threaded programming
languages like Node.js. 

Therefore one of my objectives while testing SocketCluster was to see how many
concurrent connections it can handle.

However this should not be a problem for this library, contemporary objective is
rather to achieve 10 000 000 concurrent connections like mentioned in the
excellent article in highscalability.com \citep{Reference39}. Such amount of
connections is beyond the scope of this thesis, but apparently the solution to
improve the number of connections is to move heavy lifting from the kernel to
the application itself.

Another topic to consider before begin testing was how to monitor the
application.

\section{Monitoring tool}

Monitoring tools can be divided in two categories. Basic Real time monitoring
and more convenient tool saving statics in spreadsheets and eventually
even directly plotting graphs. Most of them can be configured to record
processor load on each cores. But ideally SocketCluster tests would
require to record each threads load's. This way, if run less process than
available cores are being run the exact usage of each thread can still be 
found.

For this reason and also to have more freedom on how data is being
processed, out of the box tool have been cast aside for more basic tools like top 
and htop. htop has been used to visualize data in real time and check if the 
test was running flawlessly. top has been used in batched mode to output
the data in files.

In a latter phase, bash operation is used to format the raw data extracted from 
top's file. And finally, graphs are plotted with gnuplot.

\chapter{Experiment} 
\label{Chapter3} 
\lhead{Chapter 3. \emph{Experiment}} 

This chapter is a comprehensive benchmark of SocketCluster. to begin with, the
scalability of the client code will be checked with a client throughout test.
Then once the the client has been proof checked, the first experiment will
compare SocketCluster and engine.io. 

Then the experiment will purely focus on SocketCluster. The first one will
evaluate the influence of adding more cores on the performances. The second one
will study the influence of external parameters like the period of pings, the
size of the messages and the number of communications. And to finish a
concurrent experiment will be carried out in order to have an idea of
SocketCluster's behavior in highly parallel environment.  

\section{Client throughout}

This first section is composed of two experiment to check the client code is
behaving like expected.

\subsection{Client scalability}

SocketCluster-client makes the instantiation of a WebSocket clients on one core
quite straightforward. To deploy it on all available nodes, node.js
\texttt{fork()} function is used. A client code example is given in appendix
\ref{fig:WS_client_simplePing}.

The first experiment is a safety test. It checks if \texttt{fork()} distributes
evenly the work among the cores.

\begin{center}
  \begin{tabular}{ | l | l |}
  \hline
  \multicolumn{2}{|c|}{Parameters} \\
  \hline
    Instance type &  amazon s3 m3.2xlarge\\ 
    Experiment time & 120 s \\
    Number of new communication created at each iteration & 15 \\
    Client creation period & 1 s \\
    Type of ping & random number \\ 
    Ping period & 2.5 s \\ 
  \hline
  \end{tabular}
\end{center}

\begin{figure}[H]
	\centering
		\includegraphics[width=\textwidth]{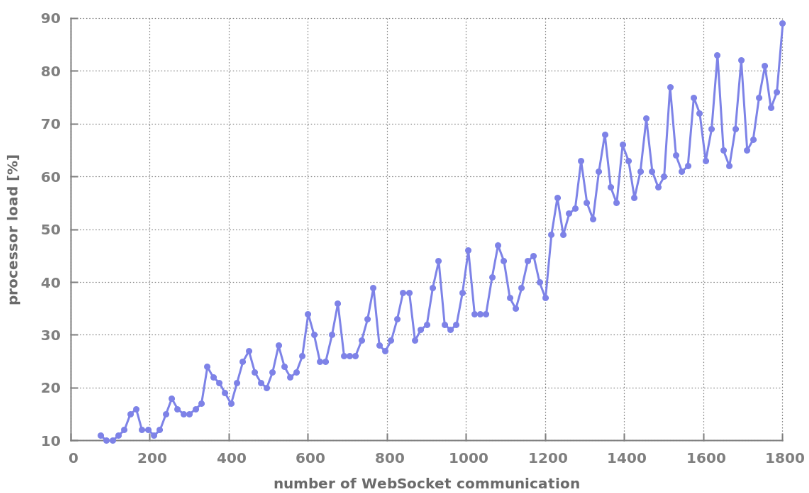}
		\includegraphics[width=\textwidth]{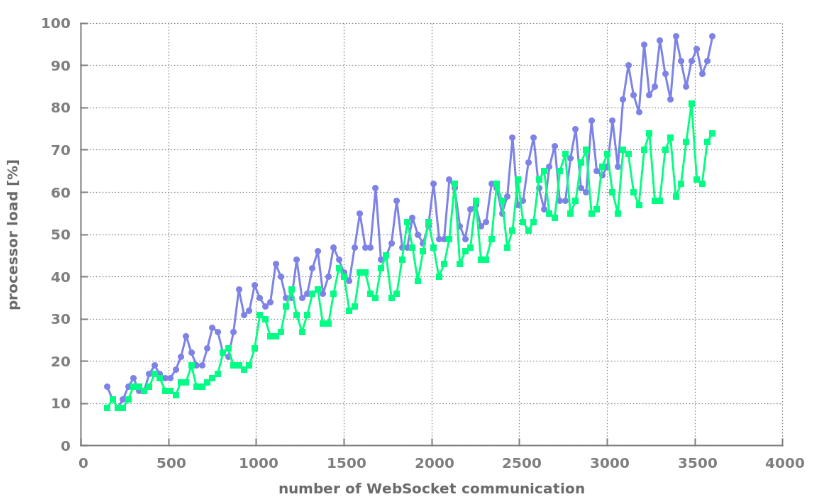}
	\caption[Client throughout]{Client throughout}
	\label{fig:1+2_client}
\end{figure}

From Figure \ref{fig:1+2_client} it can be inferred that the client
implementation works flawlessly. Adding a second core enables twice as much
communication  to be established.

\subsection{browser testing}

As mentioned in Appendix \ref{fig:index_script}, by operating minor changes in
the \texttt{index.html} file, the browser can be configured to display in real
time the number of pings received by a particular worker. If the experiment is
running locally, typing \texttt{localhost:8080} in the url will link the
browser to one worker.

\begin{figure}[H]
	\centering
		\includegraphics[width=0.9\textwidth]{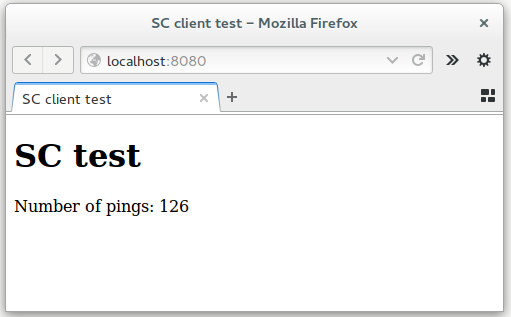}
	\caption[Browser connection to SocketCluster]{Browser connection to SocketCluster}
	\label{fig:browser}
\end{figure}

By doing so we can embody a user connected to our WebSocket server and have a
better idea of the reactivity of the server.

% --------------
% second section
% --------------

\section{Comparison with engine.io}

SocketCluster has been created to ease the creation of multi-core WebSocket
server. Logically the first experiment carried out on the server was to compare
a WebSocket to a traditional Engine.io server. 

Engine.io and SocketCluster codes can be found in Appendix \ref{SocketCluster} and \ref{engine}. 

\begin{center}
  \begin{tabular}{ | l | l |}
  \hline
  \multicolumn{2}{|c|}{Parameters} \\
  \hline
    Instance type &  amazon ec2 m3.2xlarge\\ 
    Experiment time & 60 s \\
    Number of new communication created at each iteration & 20 \\
    Client creation period & 1 s \\
    Type of ping & random number \\ 
    Ping period & 2.5 s \\ 
    Number of clients & 2 \\
  \hline
  \end{tabular}
\end{center}

\textbf{SocketCluster implementation}

\begin{figure}[H]
	\centering
		\includegraphics[width=.9\textwidth]{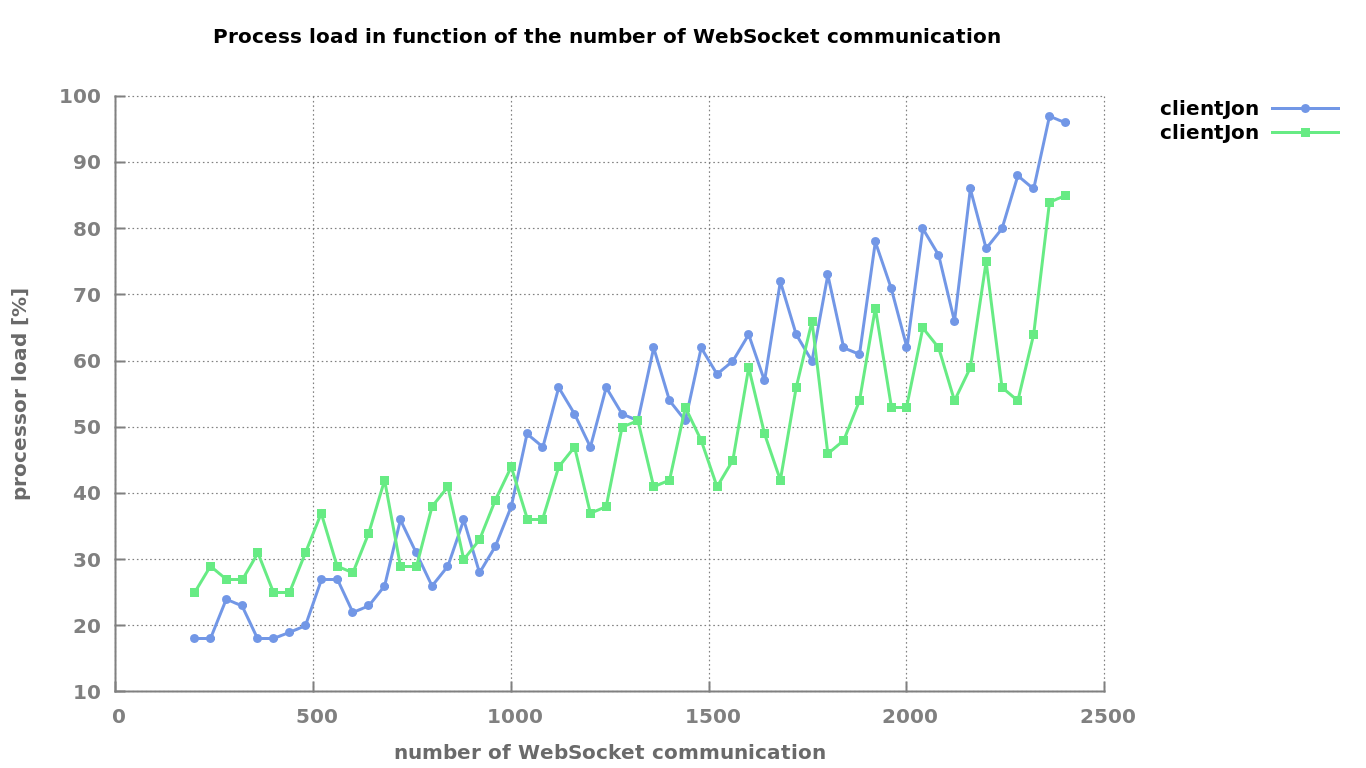}
		\includegraphics[width=.9\textwidth]{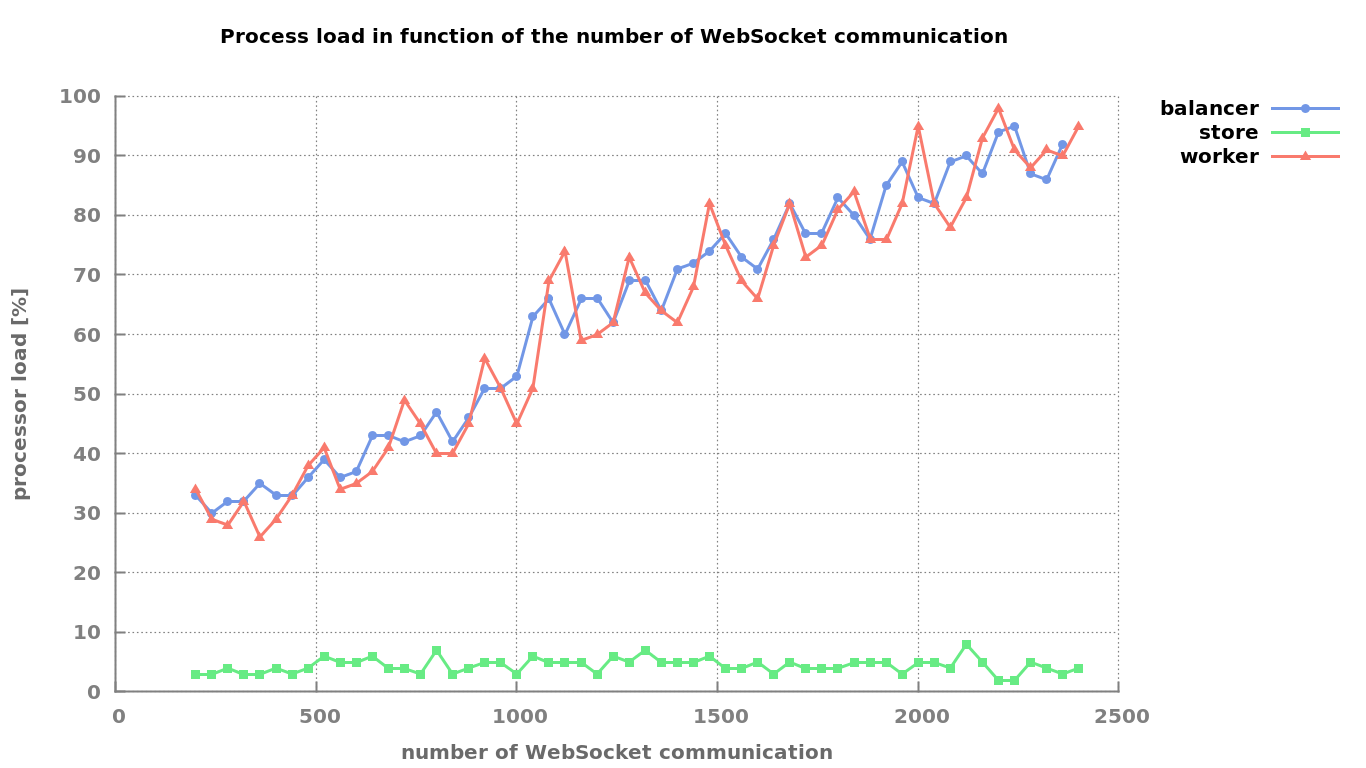}
	\caption[WebSocket implementation]{WebSocket implementation}
	\label{fig:WS_comparaison}
\end{figure}

In this experience, two clients are used to achieve a maximum of 2400 WebSocket
communications.  The server was configured to use one storage, one load
balancer and one worker. While the store processor is quite idle, the two other
processors on the other hand are almost used at full capacity.

\textbf{Engine.io implementation}
\begin{figure}[H]
	\centering
		\includegraphics[width=\textwidth]{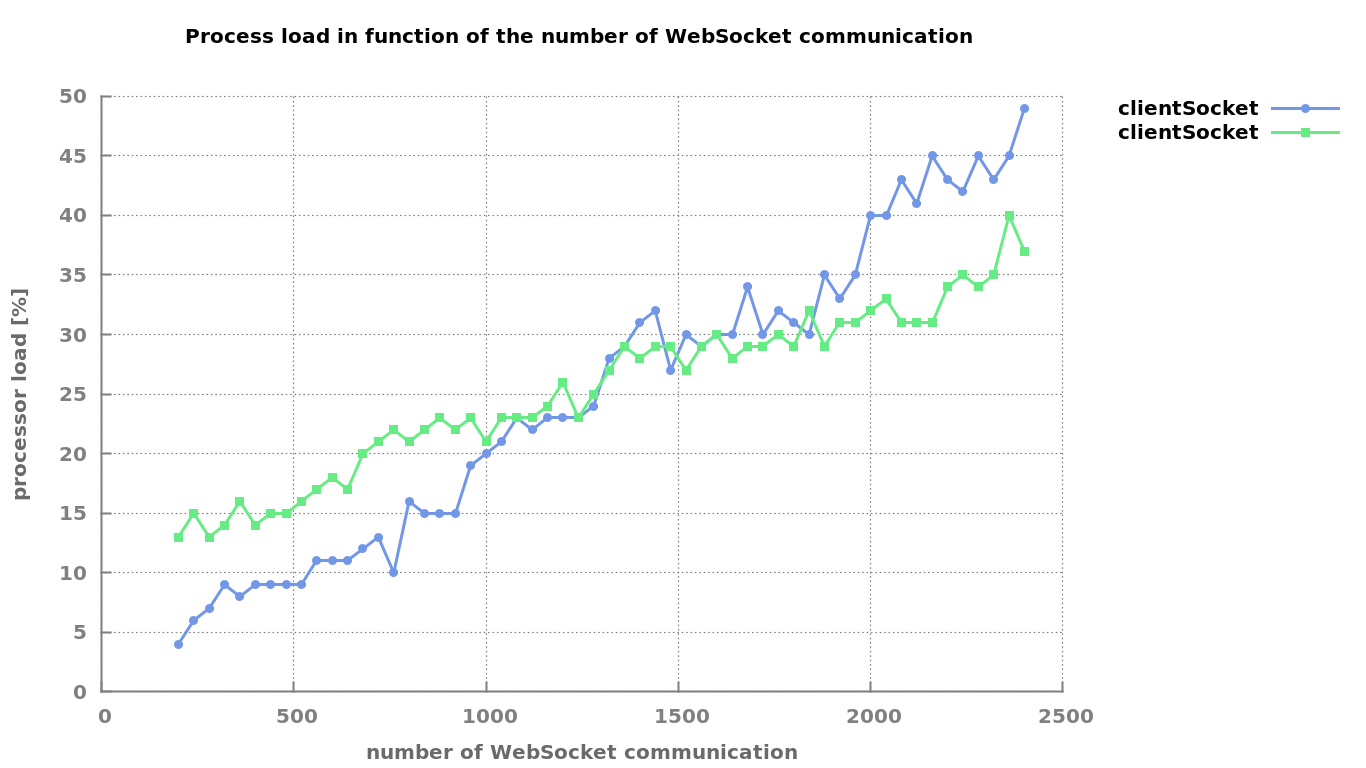}
		\includegraphics[width=\textwidth]{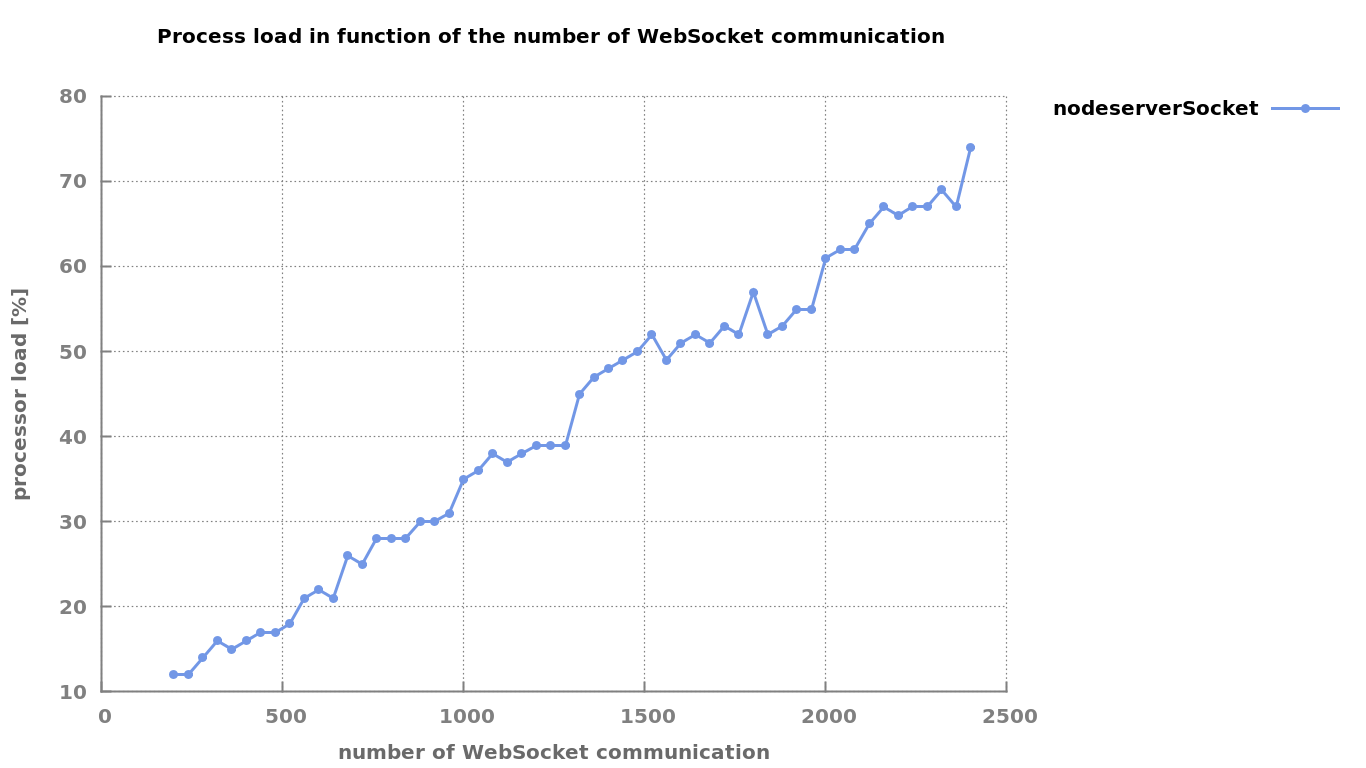}
	\caption[Engine.io implementation]{Engine.io implementation}
	\label{fig:engine_comparaison}
\end{figure}

Surprisingly, pure engine.io implementation seems to be more efficient. Clients
are hitting a maximum of 50\% processor usage compared to 90\% for WebSockets.

On the server side, engine.io processor peaks at 75\% compared to almost
100\% for WebSockets. Also even if both code have been deployed on similar
virtual machines: \texttt{amazon ec2 m3.2xlarge} the engine.io server is
running only on one core compared to three for SocketCluster (one storage,
one load balancer and one worker). This seems to show, SocketCluster is not
adapted to low number of communication.

An interesting study worth doing at this point, is to try to use
SocketCluster on one core.

% -------------
% Third section
% -------------

\section{SocketCluster context switching}

For this experiment a single core  virtual machine is used for the server:
\texttt{amazon ec2 m3.medium}.

\begin{center}
  \begin{tabular}{ | l | l |}
  \hline
  \multicolumn{2}{|c|}{Parameters} \\
  \hline
    Server instance type &  amazon ec2 m3.medium\\ 
    Client instance type &  amazon ec2 m3.2xlarge\\
    Experiment time & 80 s \\
    Number of new communication created at each iteration & 40 \\
    Client creation period & 1 s \\
    Type of ping & random number \\ 
    Ping period & 2.5 s \\ 
    Number of clients & 2 \\
  \hline
  \end{tabular}
\end{center}

\begin{figure}[H]
	\centering
		\includegraphics[width=\textwidth]{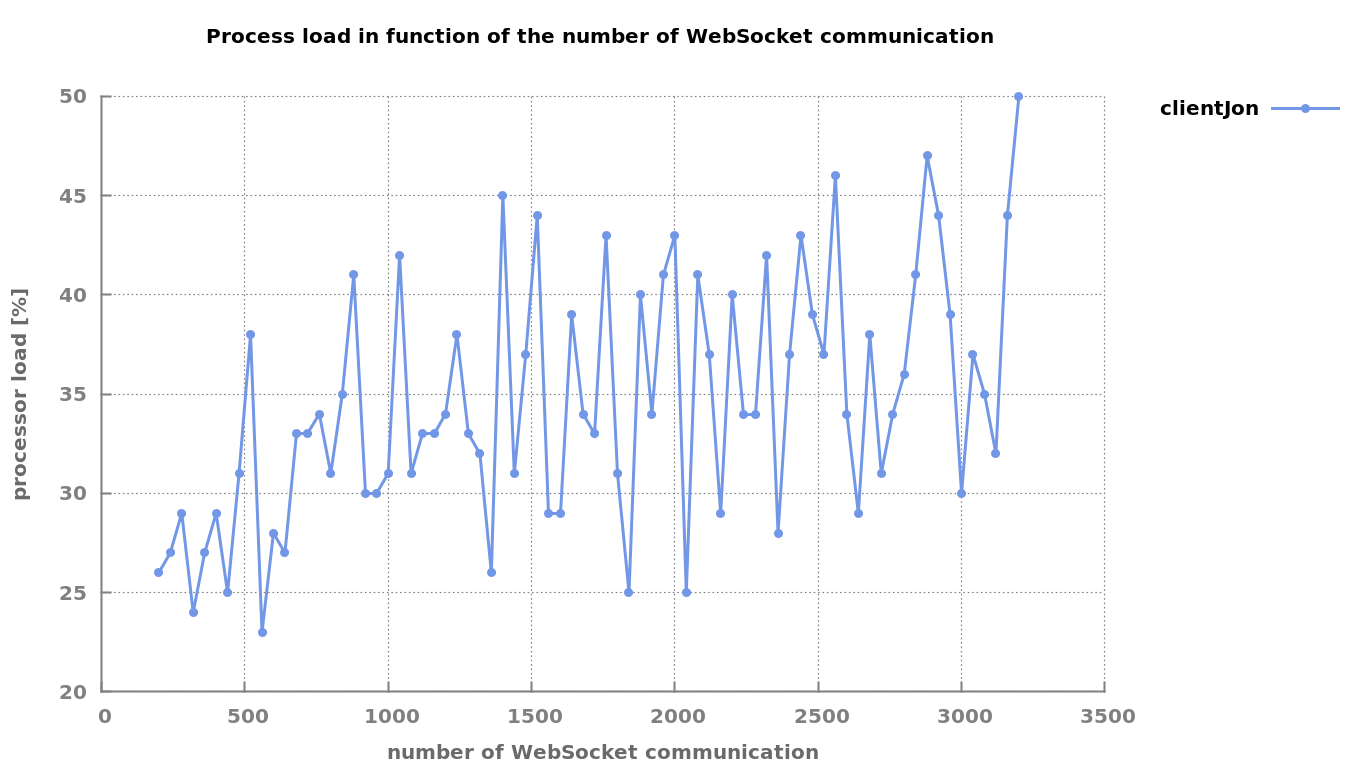}
		\includegraphics[width=\textwidth]{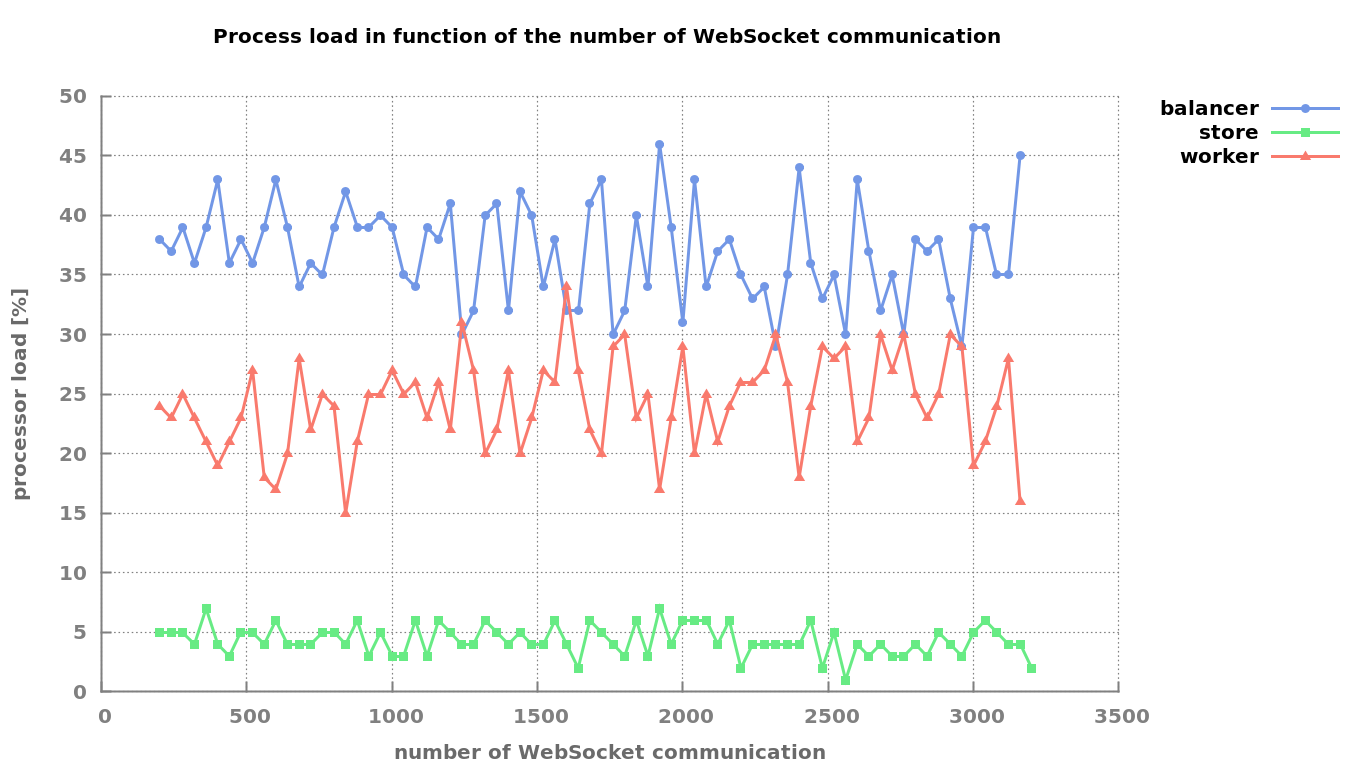}
	\caption[Context switching]{Context switching}
	\label{fig:context}
\end{figure}

At first glimpse, anyone can immediately tell there is a problem with the server
graph. The Load seems to vary randomly at an average of 40\%. What really happens, is
that most WebSocket connections are dropped shortly after being created or they 
not are even created. The problem is a single core needs to handle four threads. So each
time another application is called the context changes. The result is even worse in the case of 
a multi-processor server, because threads are then balanced between processors. Threads 
are heavy weight units, moving them introduces consequent overheads.

In conclusion, this experiment proves SocketCluster is not aimed to be used with
project which involve more threads than available cores.

% --------------
% Fourth section
% --------------

\section{Horizontal scaling of SocketCluster }

This section evaluates the performances of SocketCluster for a growing number
of processors. 

\textbf{Client code}

The client code used in all this part is the same. Two clients are used to
produce a maximum of 2400 WebSocket communications.

\begin{center}
  \begin{tabular}{ | l | l |}
  \hline
  \multicolumn{2}{|c|}{Parameters} \\
  \hline
    Instance type &  amazon ec2 m3.2xlarge\\ 
    Experiment time & 60 s \\
    Number of new communication created at each iteration & 20 \\
    Client creation period & 1 s \\
    Type of ping & random number \\ 
    Ping period & 2.5 s \\ 
    Number of clients & 2 \\
  \hline
  \end{tabular}
\end{center}

\begin{figure}[H]
	\centering
		\includegraphics[width=\textwidth]{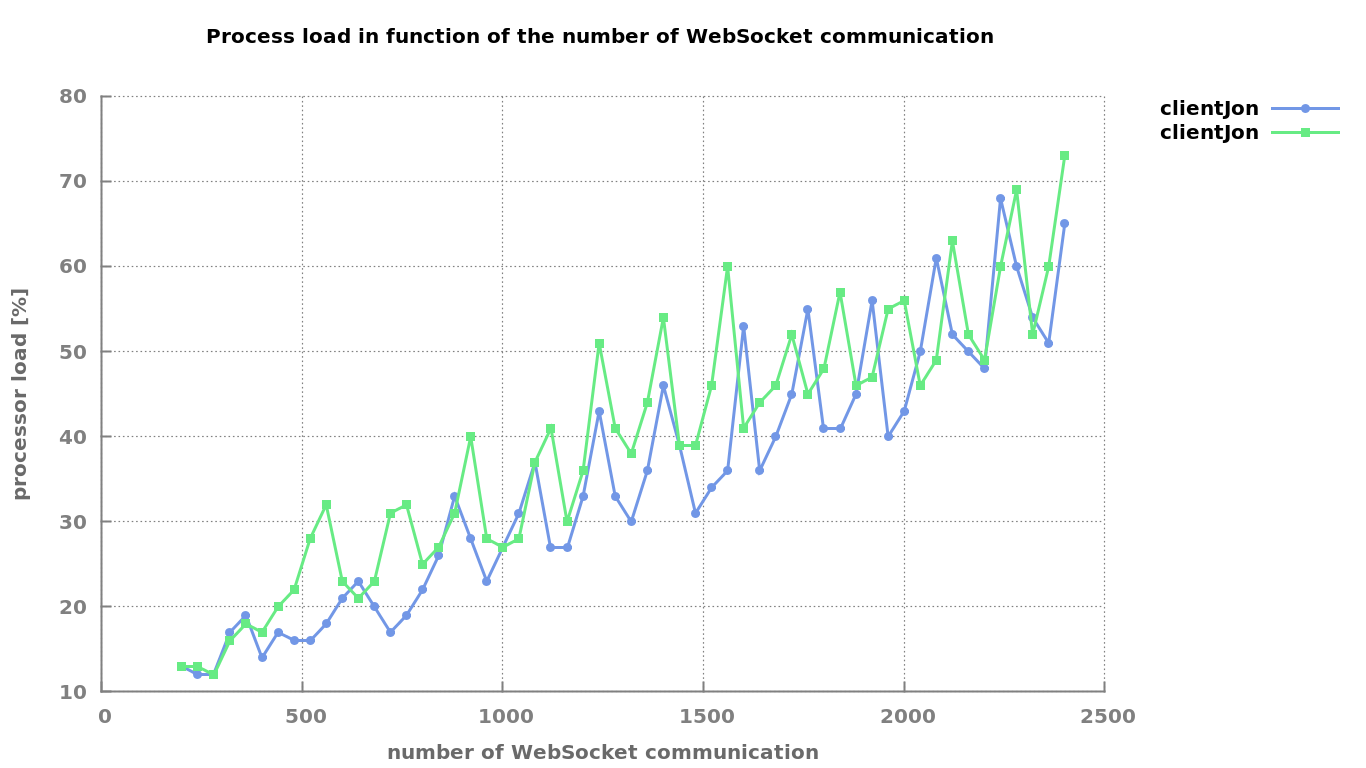}
	\caption[Simple WebSocket client]{client code}
	\label{fig:WS_client_rising}
\end{figure}

\textbf{Experiment on three cores}

The first test is run a server using a one store, one load balancer and one
worker.

\begin{figure}[H]
	\centering
		\includegraphics[width=\textwidth]{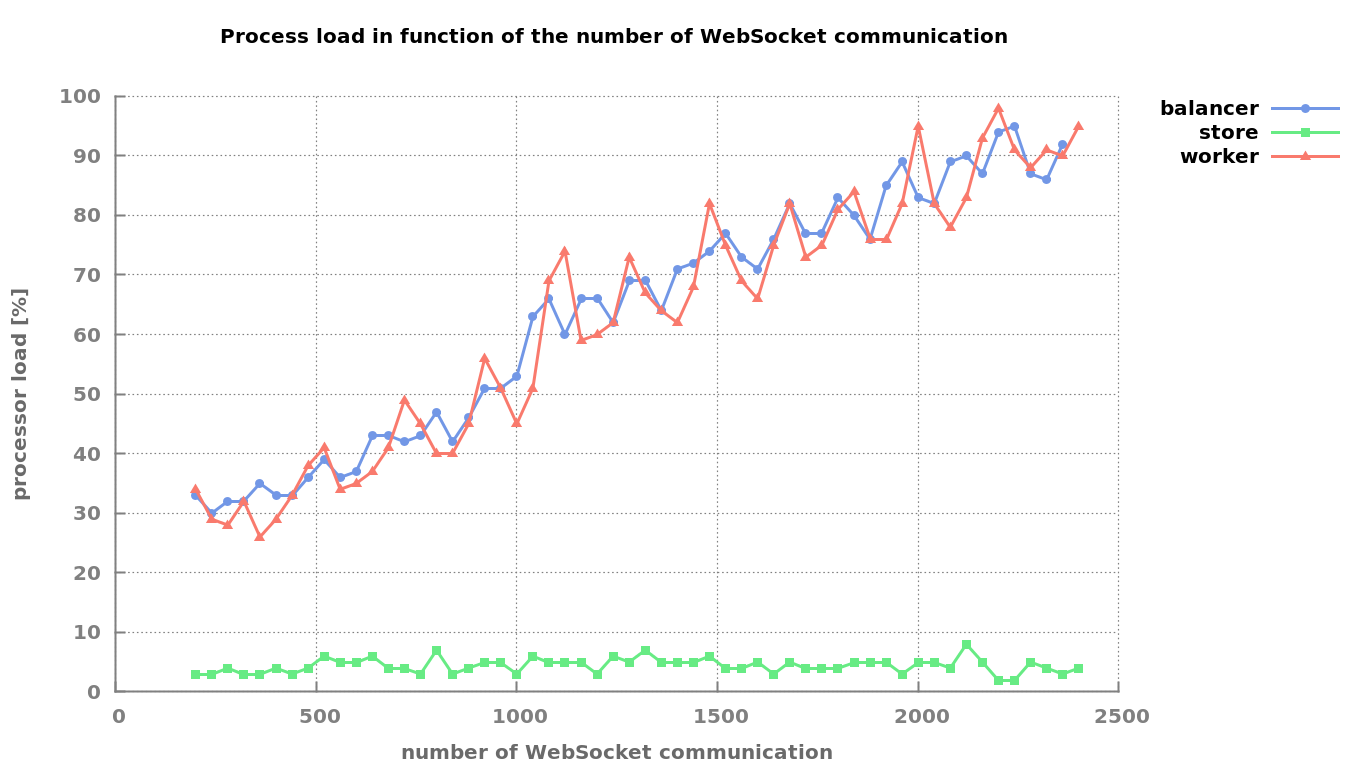}
	\caption[WebSocket server on three cores]{Server with three cores}
	\label{fig:WS_server_1rising}
\end{figure}

Figure \ref{fig:WS_server_1rising} clearly shows the worker and load balancer
cores are almost used to their full extent. In order to handle more communication more
cores should be added. 

\textbf{Experiment on five cores}

\begin{figure}[H]
	\centering
		\includegraphics[width=.9\textwidth]{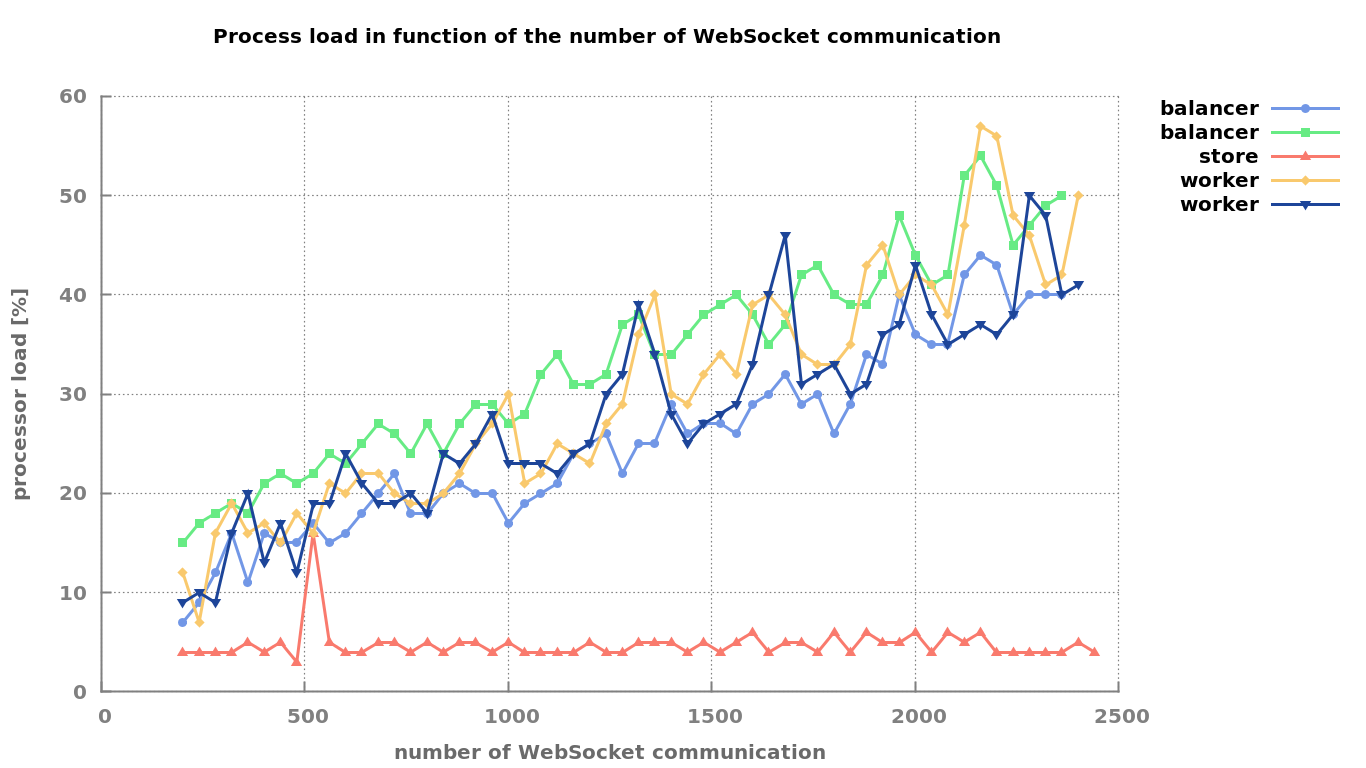}
	\caption[WebSocket server on five cores]{server with five cores}
	\label{fig:WS_server_2rising}
\end{figure}

In this experiment two more cores have been added. Load balancers and
workers nicely balance the work between themselves and the maximum load drops to 50\%.\\

\textbf{Experiment on seven cores}

\begin{figure}[H]
	\centering
		\includegraphics[width=\textwidth]{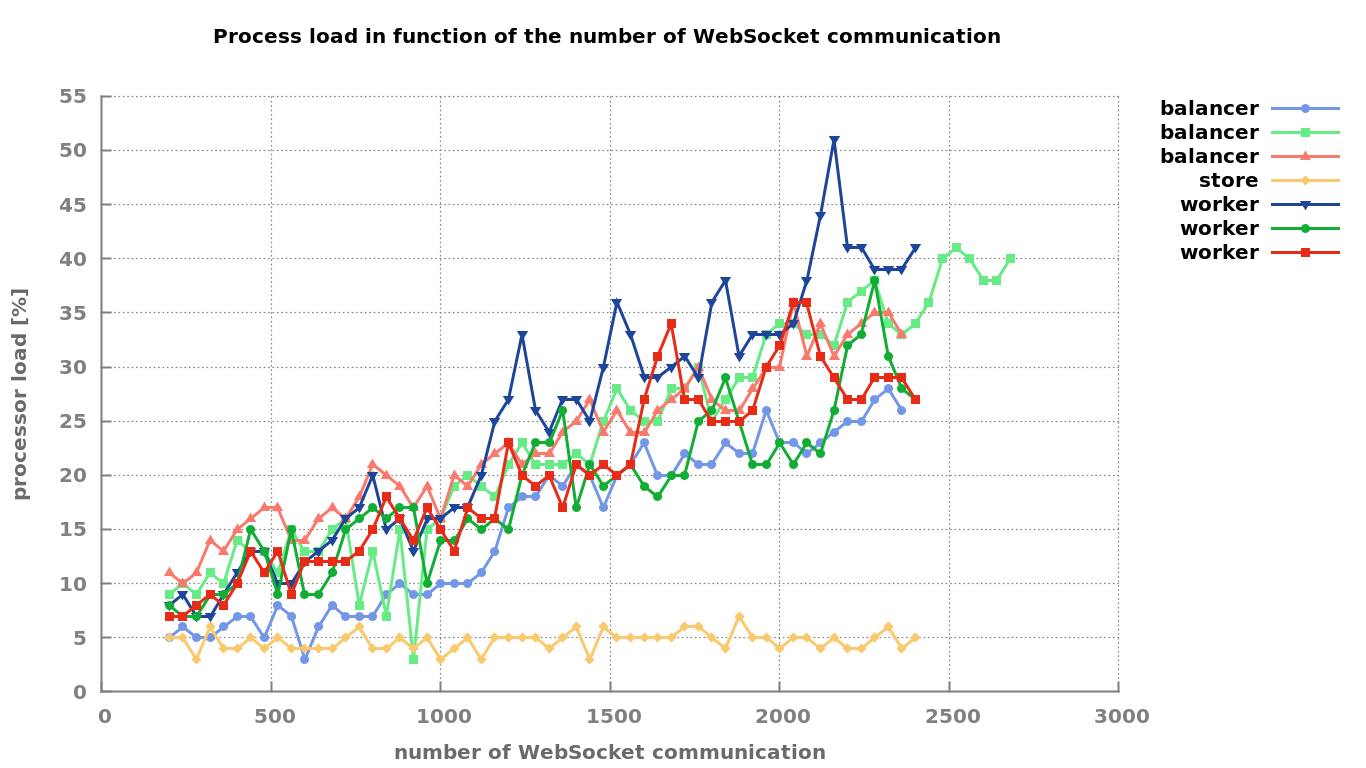}
	\caption[WebSocket server on seven cores]{server with seven cores}
	\label{fig:WS_server_3rising}
\end{figure}

This last test is less conclusive. With a total of 3 cores for load balancers
and three for workers the processors load varies between 30\% and 50\%
depending on the task.

As expected, in the long run by adding more processor SocketCluster's
performance get better then engine.io. However in case n is the number of
available processor, SocketCluster is not n times more effective then
engine.io.

\newpage

Actually in this experiment it seems that an equivalent number of workers and
load balancers are needed for the application to run seamlessly. In case the
application doesn't use a store, to gain twice as much computational power, twice
as many processor are required. This makes SocketCluster $\frac{n}{2}$ time
more efficient then engine.io.

Furthermore, it showed adding too many cores is a waste of resources. This
stresses the importance of finding a load balancer/worker/store ratio rule.

% -------------
% Fifth section
% -------------
\section{Parameters' influence}

This section aims at determining which parameter between the number of
WebSocket communications, the period of the pings and the size of the message
exchanged has the most influence on the server processor usage. 

The library \texttt{delivery} has been used to transfer file over WebSocket.
The code can be found in Appendix \ref{fig:WS_server_delivery}.

\begin{center}
  \begin{tabular}{ | l | l |}
  \hline
  \multicolumn{2}{|c|}{Fixed parameters} \\
  \hline
    Instance type &  amazon ec2 m3.2xlarge\\ 
    Experiment time & 60 s \\
    Number of new communication created at each iteration & 20 \\
    Client creation period & 1 s \\
    Type of ping & random number \\ 
  \hline
  \end{tabular}
\end{center}

\newpage

\textbf{Reference experiment}

This first experience will be taken as a reference for the next ones. It has
been carried out with 2 clients establishing together a total of 2400
communications. The period of the pings is in average four seconds and the
size of the file exchanged is 81 bytes.

\begin{figure}[H]
	\centering
		\includegraphics[width=\textwidth]{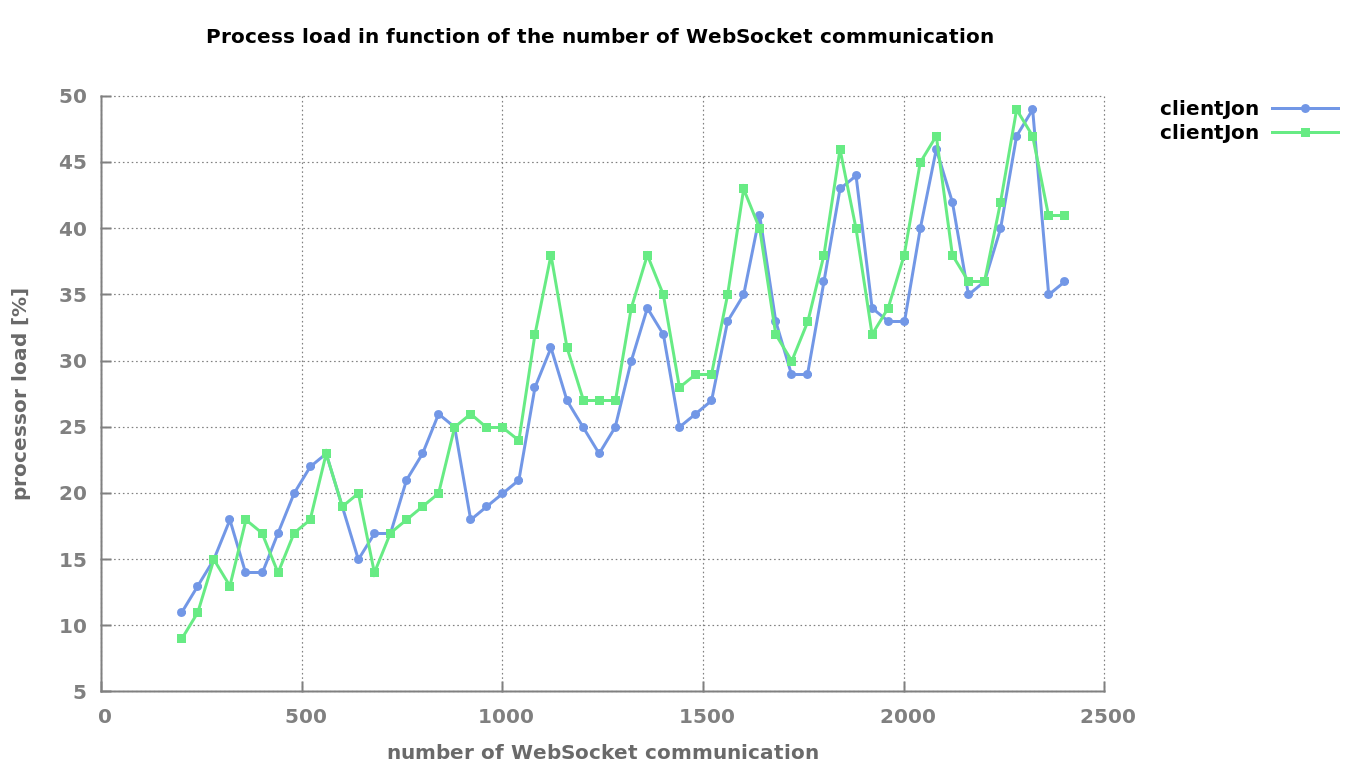}
		\includegraphics[width=\textwidth]{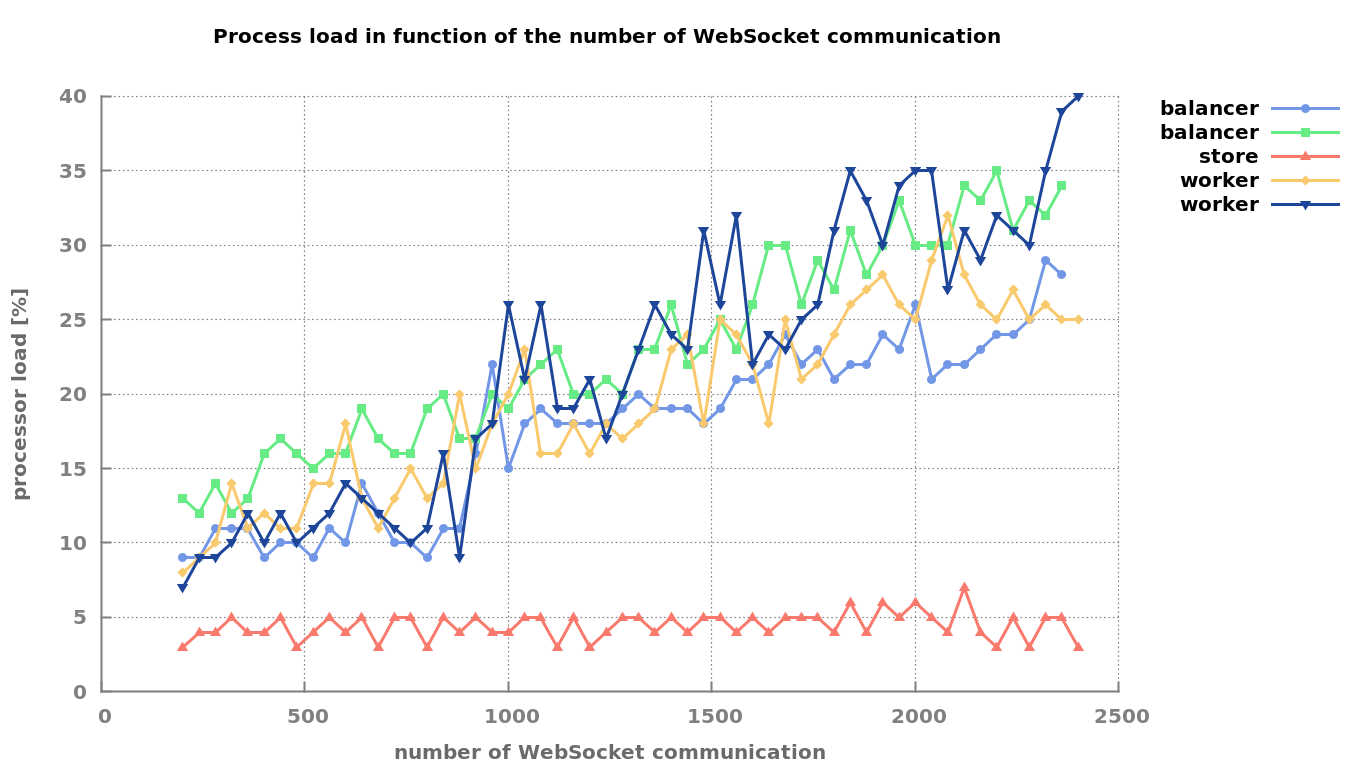}
	\caption[Reference experiment]{Reference experiment}
	\label{fig:base_influence}
\end{figure}

\newpage

\textbf{Pings' period experiment}

In this experiment, the average time separating two pings has been decreased
from four to three seconds.

\begin{figure}[H]
	\centering
		\includegraphics[width=.9\textwidth]{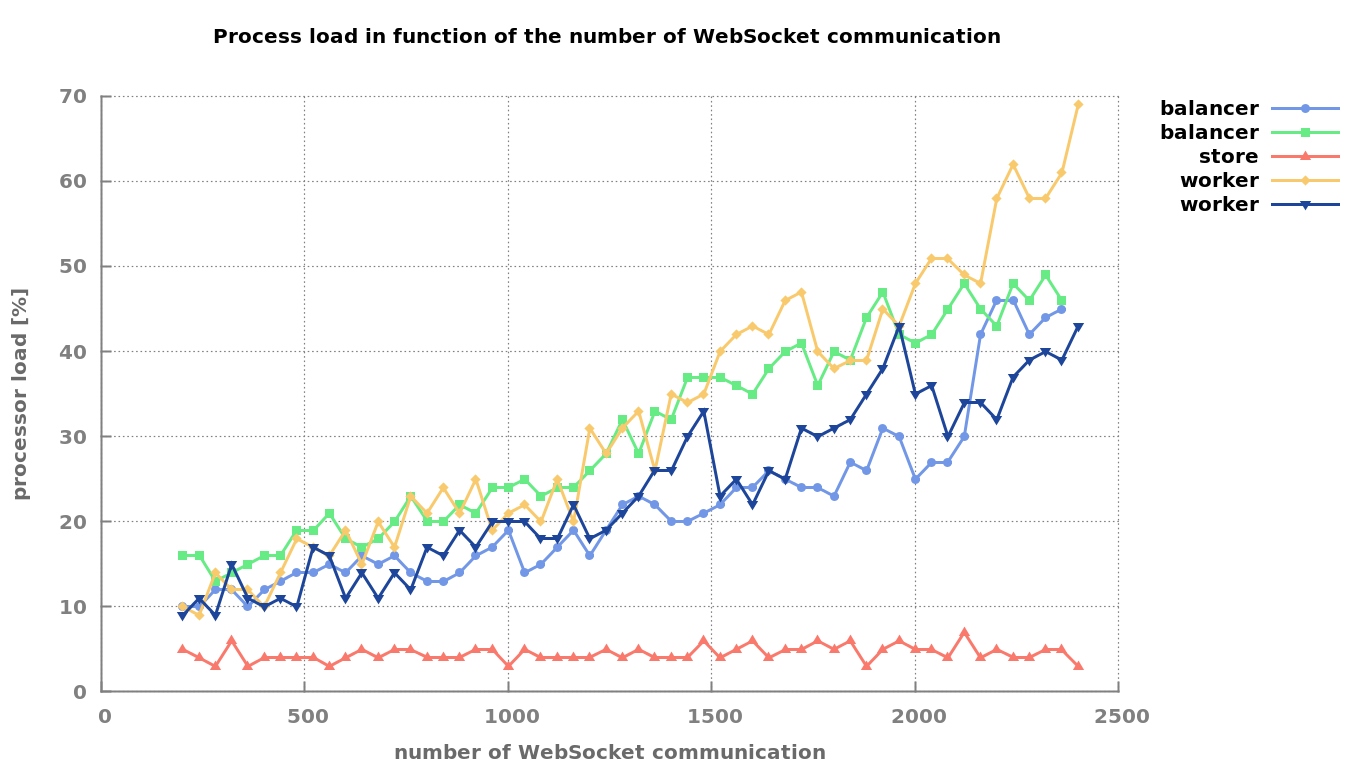}
	\caption[Pings' period experiment]{Pings' period experiment}
	\label{fig:ping_server_influence}
\end{figure}

\textbf{Amount of WebSocket communication experiment}

The following Figure stresses the influence of the number of WebSocket communication
channel. To obtain more communication, a third client has been added compared to 
the reference experience \ref{fig:base_influence}.

\begin{figure}[H]
	\centering
		\includegraphics[width=.9\textwidth]{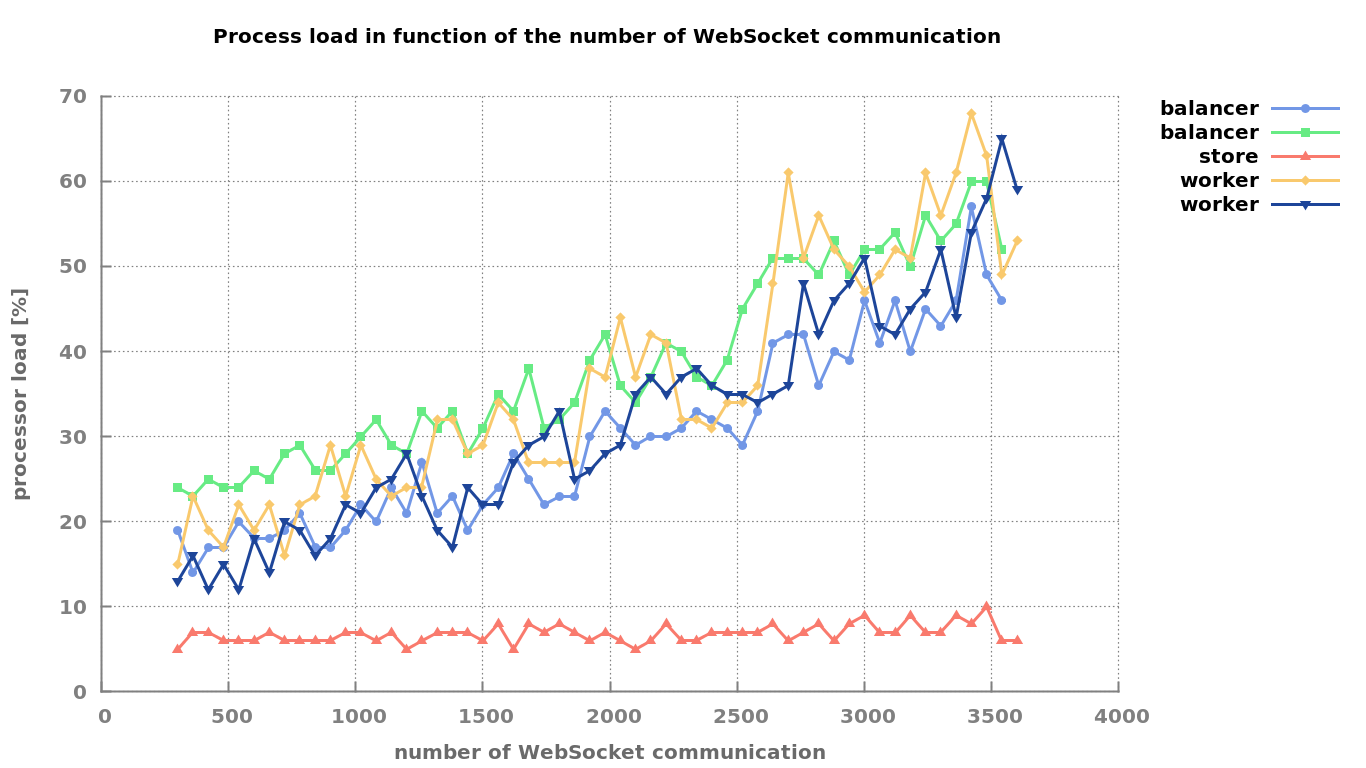}
	\caption[Amount of WebSocket communication experiment]{Amount of WebSocket communication experiment}
	\label{fig:communication_server_influence}
\end{figure}

\textbf{Size of exchanged files experiment}

This last graph underlines the influence of the size of files. The file transferred in this experiment is
500 kbytes compared to 1 kbytes for the reference experience.

\begin{figure}[H]
	\centering
		\includegraphics[width=\textwidth]{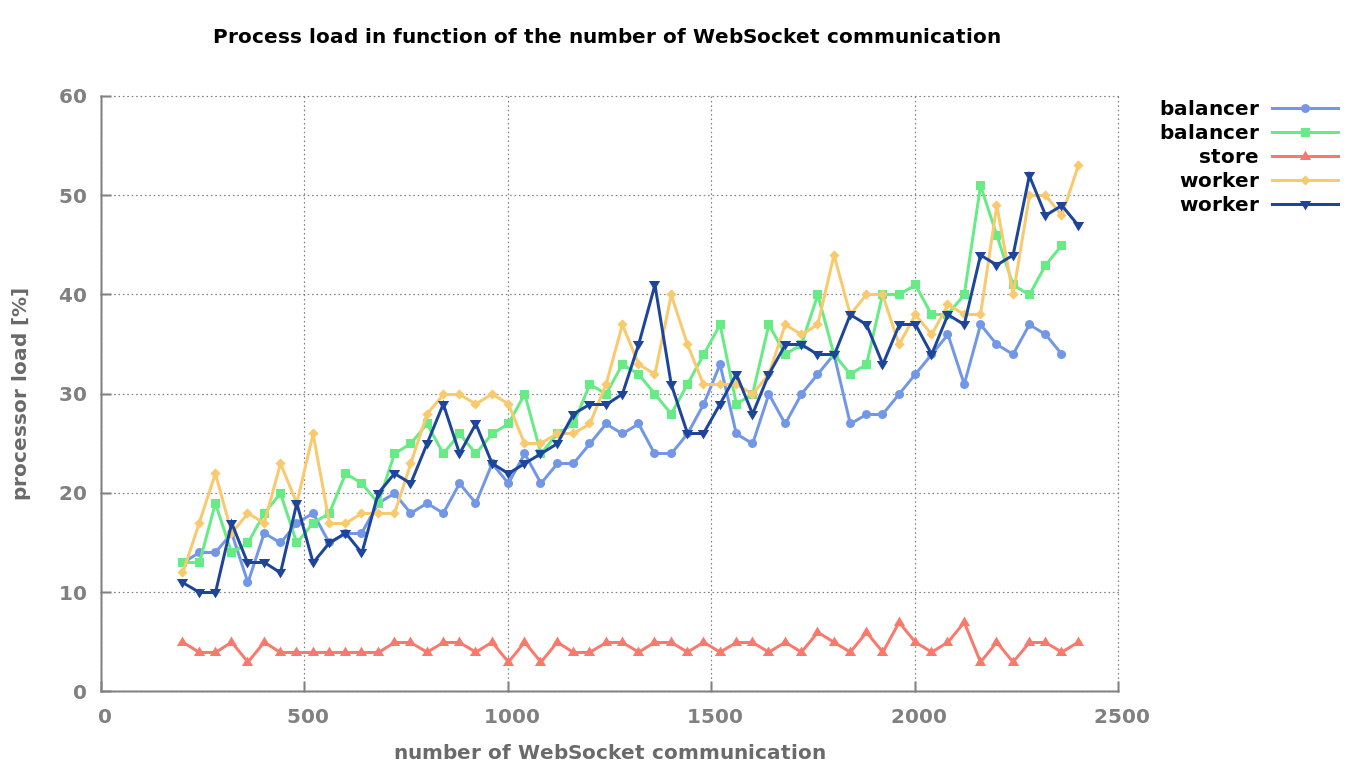}
	\caption[File size experiment]{File size experiment}
	\label{fig:file_server_influence}
\end{figure}

In conclusion, it seems that the size of the exchanged files isn't as important
as the rate of pings and the number of WebSockets communications.

% -------------
% sixth section
% -------------

\section{Concurrent connections experiment}

This study was done to investigate the number of connections a
single server can handle. As seen in the previous section, the number of
connections is tightly bound to the parameters used to simulated the clients
interactions with the server.  Lets suppose each client receives a small file
from the server every 2.5 seconds in average.

\begin{center}
  \begin{tabular}{ | l | l |}
  \hline
  \multicolumn{2}{|c|}{Parameters} \\
  \hline
    Server instance type &  amazon ec2 c3.8xlarge\\ 
    Client instance type & amazon ec2 c3.4xlarge\\
    Experiment time & 150 s \\
    Number of new communication created at each iteration & 10 \\
    Client creation period & 1 s \\
    Ping period & 6 s \\ 
    Size of the file exchanged & small \\
  \hline
  \end{tabular}
\end{center}

\begin{figure}[H]
	\centering
		\includegraphics[width=\textwidth]{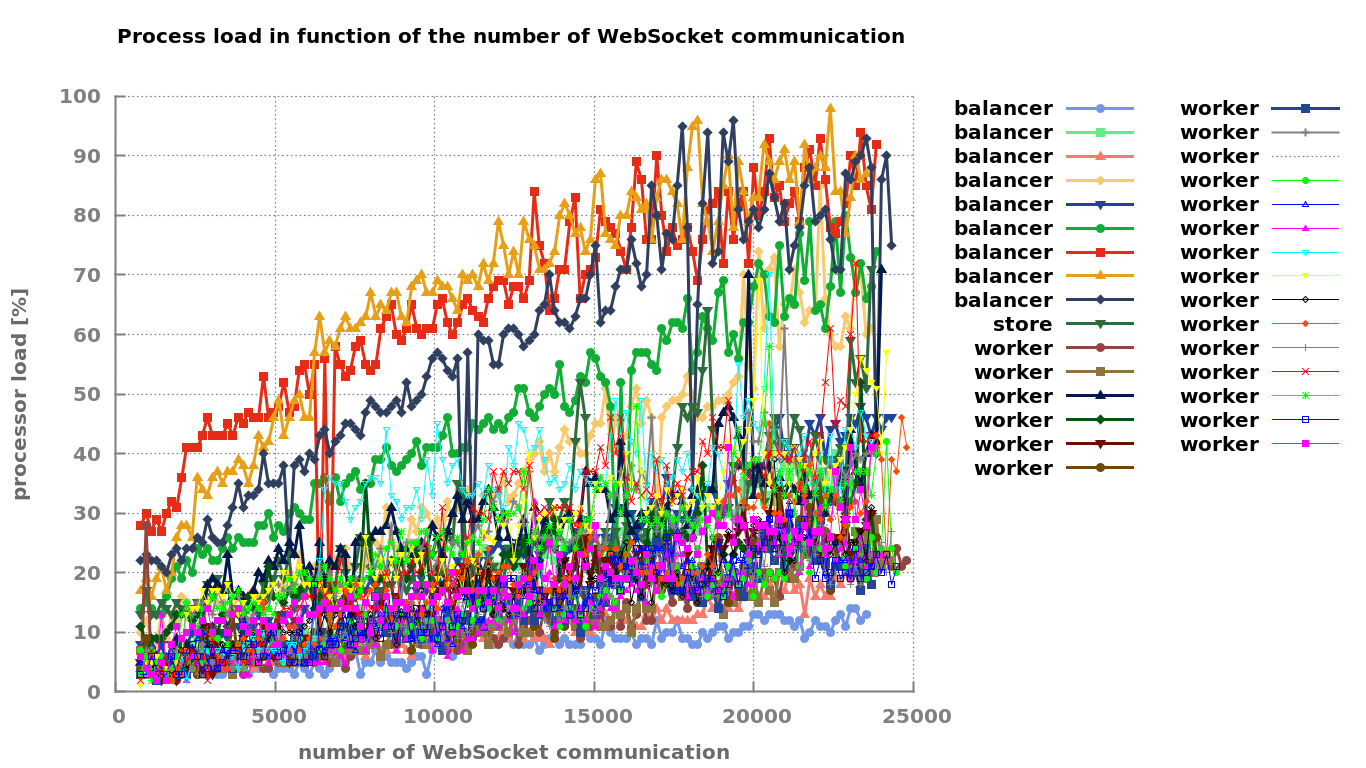}
		\includegraphics[width=\textwidth]{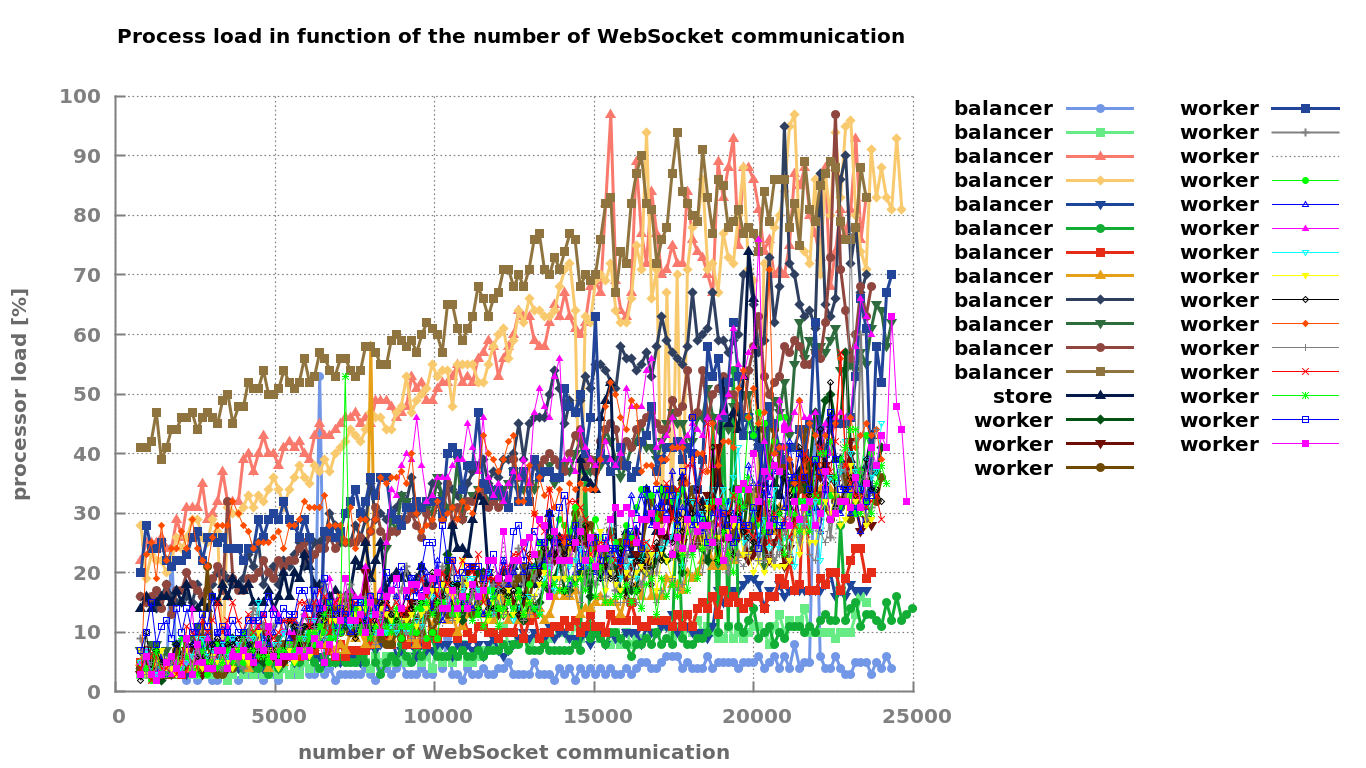}
	\caption[Maximum number of WebSocket communication]{Maximum number of WebSocket communication}
	\label{fig:max_communication}
\end{figure}

This experiment has been carried out on the biggest server made available by
amazon ec2.  SocketCluster will hardly be used in this conditions during normal
usages.  It is way cheaper to make clusters of SocketClusters server then to
use a beastly server like this one.

This experiment confirmed SocketCluster is able to support around 25 000
concurrent connections. It also pointed out an imperfection of SocketCluster.
The first graph is an experiment with 9 load balancers and 21 workers. The
second with 12 load balancers and 18 workers.  In the first experiment the load
balancers' load is quite high. Adding more communication will result in
communication to be dropped, as a result the second experiment has been carried
out with more load balancers. But the previous figures clearly show that some
load balancer are still using way too much computing power and some on the
other hand are almost idle.

This points out a bad load balancing between the load balancers themselves.

\section{Experiment summary}

The client throughout tests showed the number of communications are scaling
linearly when adding more cores. It also gave an insight into the user
experience when using SocketCluster.

The second section was a little disappointing, one would expect a SocketCluster
code running on three cores to achieve better then a regular engine.io code
running on one core. However it is not the case, engine.io is significantly
better.

The third section stresses the importance of running SocketCluster on a less
processus then available cores. Otherwise the operating system as to operate
heavy weight context switching.

The forth section studies the horizontal scaling of SocketCluster. Apparently,
in a relatively low parallel environment, the application needs as much
load-balancers as worker. And once they get saturated, adding a load-balancer
and a worker will efficiently increase the performances.

The fifth experiment demonstrated the number of communication and the periods
of pings increase the processor usage more quickly then the size of the
messages exchanged.

The last experiment which was intended as a pure concurrent experiment, proved
SocketCluster can handle 25k communications on a single server. But more
importantly it showed that in a highly parallel environment, the load balancers
begin to miss behave.

\chapter{Conclusion} 
\label{Chapter4} 
\lhead{Chapter 4. \emph{Conclusion}} 

After studying the current research around WebSocket in a distributed
environment, this thesis focused on benchmarking node.js's real time
engine SocketCluster.

SocketCluster is a promising library still actively under development.
It efficiently provides a highly scalable WebSocket server that makes
use of all available cpu cores on an instance. It removes the limitation
of having to run node.js code on single cores.

\textbf{Experiment conclusion}

Experiments carried out on SocketCluster revealed two main limitations.  If
running on comparable hardware, a SocketCluster worker will be less efficient
then a basic engine.io implementation. Also SocketCluster efficiency
dramatically drops if it is run with more process then available cores because
of context switching.

SocketCluster should be used in highly parallel environment and
therefore these limitations rarely apply. SocketCluster theoretically  enables
user to scale an application vertically  without limits. N being the number of
cores the server has, SocketCluster has been proved to be at least
$\frac{N}{2}$ more efficient then a basic node.js implementation.  As the
number of cores rises, it looks like the performance could be slightly better
then $\frac{N}{2}$. The load balancers begins to misbehave and performance
is limited by a few overloaded load balancers. However, it is probably only a
question of time until a patch fixes this issue.

\newpage
\textbf{Future work}

While benchmarking SocketCluster, useful SocketCluster features were
considered.  System administrators could benefit from a real-time monitoring tool
to check the state of each threads and thus help them manage the size of the
cluster. The monitoring tool could even be linked with an algorithm to
automatically append or delete threads.  SocketCluster would then be an autonom
entity. Scaling on its own without any human interaction.

Also further studies could be carried on SocketCluster on more then on server.
Since each cores already operates as a separate thread, the perforance
shouldn't decrease if spread on many servers. But it might be worth checking.

%% ----------------------------------------------------------------
% Now begin the Appendices, including them as separate files

\addtocontents{toc}{\vspace{2em}} % Add a gap in the Contents, for aesthetics

\appendix % Cue to tell LaTeX that the following 'chapters' are Appendices

\chapter{SocketCluster}
\label{SocketCluster}
\lhead{Appendix A. \emph{SocketCluster}}
\section{Simple ping-pong exchange}
\textbf{Client code}

This is an example of a WebSocket client code spread on all available cores.
New clients are spawned every \texttt{numberClientsEachSecond}. Thereafter,
every \texttt{intv} each clients sends a ping event cast to a Javascript JSON
object.

\begin{figure}[H]
	\centering
		\includegraphics[width=0.7\textwidth]{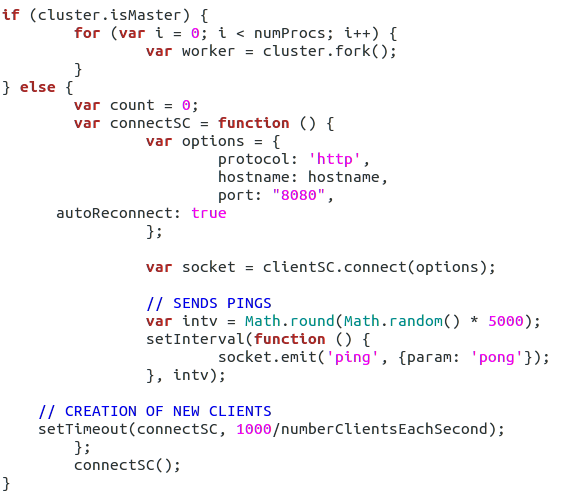}
	\caption[Simple WebSocket client code]{Pings from client}
	\label{fig:WS_client_simplePing}
\end{figure}

To best simulate clients interaction with a websocket server, new sockets are
created at random intervals \texttt{intv = Math.round(Math.random()*5000)}.

\textbf{Server code}

The server listens for pings event and answers back with pongs event. In this
case the pong event is an integer counting the number of pings this particular
worker had during the whole experiment.

\begin{figure}[H]
	\centering
    \includegraphics[width=\textwidth]{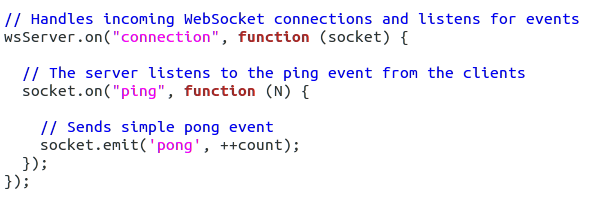}
	\caption[Simple WebSocket server code]{Server answering with pongs}
	\label{fig:WS_server_simplePong}
\end{figure}

\section{File transfer}

\textbf{Client code}

In this example, the goal is to exchange a file using the WebSocket protocol.
For this purpose, the node.js \texttt{delivery} library is used.

New clients are created on the same model as the previous example. Each new
client is stored in the \texttt{clients} array. Each clients are also
periodically sending pings. The only add on is the \texttt{map} function to
enable the each socket to retrieve the document sent by the server. 

\begin{figure}[H] \centering
  \includegraphics[width=\textwidth]{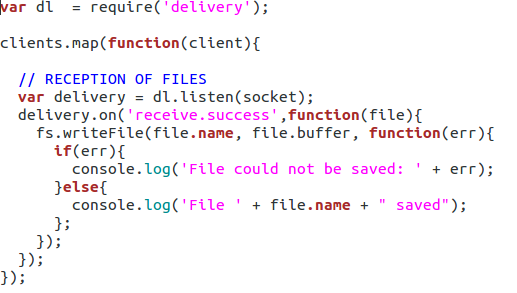}
\caption[Client code for file transfers with WebSocket ]{Clients receptionning files} 
\label{fig:WS_client_delivery}
\end{figure}

\textbf{Server code}

The server listens for pings. And sends back a file, \texttt{foo.txt} in this
example.

\begin{figure}[H]
	\centering
    \includegraphics[width=\textwidth]{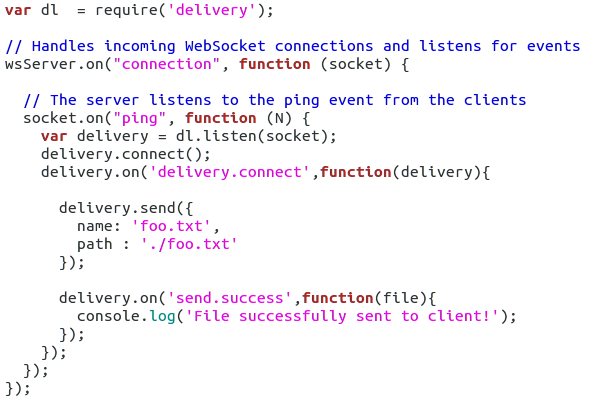}
	\caption[Server code for file transfers with Websocket]{Server sending files}
	\label{fig:WS_server_delivery}
\end{figure}

\chapter{Engine.io}
\label{engine}
% \lhead{Appendix A. \emph{SocketCluster}}

This appendix gives the code used to create a simple engine.io server and
client. Comparison with SocketCluster code in appendix \ref{SocketCluster} 
shows the difference between both implementation is small. 

In fairness, SocketCluster API is very close to engine.io.

\textbf{Client code}

\begin{figure}[H]
	\centering
		\includegraphics[width=0.9\textwidth]{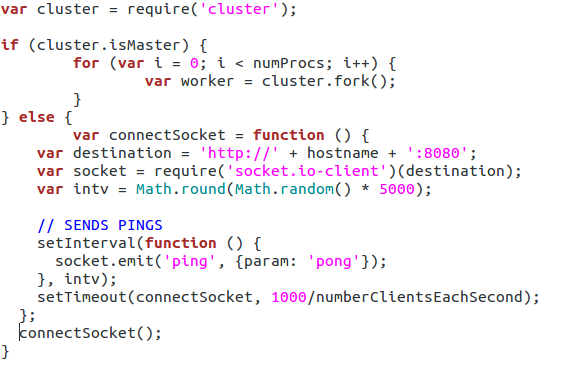}
	\caption[Engine.io client code]{Pings from client}
	\label{fig:engine_client_simplePing}
\end{figure}

\newpage

\textbf{Server code}

\begin{figure}[H]
	\centering
    \includegraphics[width=0.7\textwidth]{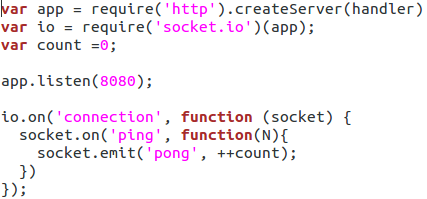}
	\caption[Engine.io server code]{Server answering with pongs}
	\label{fig:engine_server_simplePong}
\end{figure}

\chapter{Real time throughout check}
\label{indexHTML}

By inserting the following script in \texttt{index.html} the browser will
display in real-time the number of pings received by a WebSocket server.

\begin{figure}[H] \centering
  \includegraphics[width=0.8\textwidth]{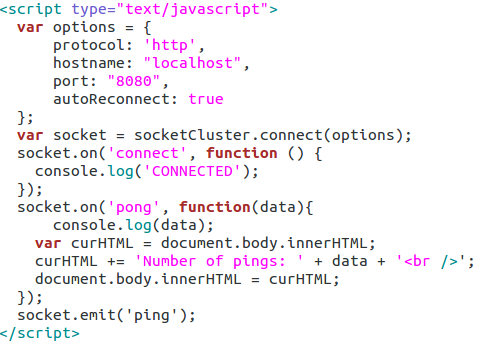}
\caption[Modification to index.html]{Modification to \texttt{index.html}}
\label{fig:index_script} \end{figure}

All it does is emitting a ping, then listening to the pong event and displaying
it directly in the html page. The pong payload as can be seen in
\ref{fig:WS_server_simplePong} is \texttt{count}, an integer incremented each
new ping.

\addtocontents{toc}{\vspace{2em}}  % Add a gap in the Contents, for aesthetics
\backmatter

%% ----------------------------------------------------------------
\label{Bibliography}
\lhead{\emph{Bibliography}}  % Change the left side page header to "Bibliography"
\bibliographystyle{unsrtnat}  % Use the "unsrtnat" BibTeX style for formatting the Bibliography
\bibliography{Bibliography}  % The references (bibliography) information are stored in the file named "Bibliography.bib"

\end{document}